%% file: hotnets26.tex
\pdfoutput=1
\PassOptionsToPackage{dvipsnames}{xcolor}
\documentclass[sigconf,10pt]{acmart}
\copyrightyear{2026}
\acmYear{2026}
\setcopyright{cc}
\setcctype{by}
\acmConference[HotNets '26]{The 25th ACM Workshop on Hot Topics in Networks}{November 16--17, 2026}{Salt Lake City, UT, USA}
\acmBooktitle{The 25th ACM Workshop on Hot Topics in Networks (HotNets '26), November 16--17, 2026, Salt Lake City, UT, USA}
\acmDOI{10.1145/3847290.3859939}
\acmISBN{979-8-4007-3029-0/2026/11}

\usepackage[utf8]{inputenc} %
\usepackage[T1]{fontenc}    %
\usepackage{hyperref}       %
\usepackage{url}            %
\usepackage{booktabs}       %
\usepackage{nicefrac}       %
\usepackage{microtype}      %
\usepackage{xcolor}         %
\usepackage{amsmath, amsthm}
\usepackage{graphicx}
\usepackage{subcaption}
\usepackage{enumitem}
\usepackage{multirow}
\usepackage{cleveref}
\usepackage[font=small,labelfont=bf]{caption}
\usepackage[most]{tcolorbox}
\usepackage[subtle,paragraphs=normal]{savetrees}
\usepackage{upquote}
\usepackage{etoc}

\input{preamble}

\title{Instantiating the Knowledge Plane}
\author{Francisco Pereira}
\affiliation{
    \institution{INESC-ID, Instituto Superior Técnico, University of Lisbon}
    \city{Lisbon}
    \country{Portugal}
}
\author{Justine Sherry}
\affiliation{
    \institution{Carnegie Mellon University}
    \city{Pittsburgh}
    \country{USA}
}
\author{Pratiksha Thaker}
\affiliation{
    \institution{Databricks}
    \city{New York}
    \country{USA}
}

\begin{document}
\etocdepthtag.toc{mainmatter}

\input{sections/abstract}

\ccsdesc[500]{Networks~Network management}
\ccsdesc[500]{Computing methodologies~Multi-agent systems}

\keywords{Knowledge Plane, Multi-Agent Systems, Interdomain Routing}

\maketitle

\input{sections/intro}
\input{sections/defining-kp}

\input{sections/case-studies}
\input{sections/discussion}

\input{sections/conclusion}
\input{sections/acknowledgments}

\bibliographystyle{ACM-Reference-Format}
\bibliography{hotnets26}

\end{document}

%% file: preamble.tex
\usepackage{xspace}

\definecolor{boxbg}{HTML}{F4F8FC}
\definecolor{boxframe}{HTML}{5D8AA8}
\definecolor{titlebg}{HTML}{73A5C6}

\crefformat{section}{\S#2#1#3}
\Crefformat{section}{\S#2#1#3}
\crefrangeformat{section}{\S#2#1#3--\S#5#4#6}
\Crefrangeformat{section}{\S#2#1#3--\S#5#4#6}
\crefmultiformat{section}{\S#2#1#3}{ and~\S#2#1#3}{, \S#2#1#3}{ and~\S#2#1#3}
\Crefmultiformat{section}{\S#2#1#3}{ and~\S#2#1#3}{, \S#2#1#3}{ and~\S#2#1#3}

\usepackage{tikz}
\usetikzlibrary{decorations.pathreplacing, arrows.meta}
\usepackage[dvipsnames]{xcolor}         %
\usepackage[most]{tcolorbox}

\newtcolorbox{myquote}{
    fontupper=\itshape,
    boxrule=0.5pt,
    arc=2pt, %
    left=2mm,
    right=2mm,
    top=2mm,
    bottom=2mm,
    breakable
}

\usepackage[most]{tcolorbox}

\tcbset{
  myquote base/.style={
    fontupper=\itshape,
    boxrule=0.5pt,
    arc=2pt,
    left=2mm,
    right=2mm,
    top=3mm,
    bottom=2mm,
    breakable,
    enhanced,
    attach boxed title to top left={xshift=2mm,yshift*=-2mm},
    coltitle=black,
    boxed title style={
      size=small,
      colframe=black,
      boxrule=0.5pt,
      arc=2pt,
    }
  },
  model tag/.style={
    overlay unbroken and first={
      \node[anchor=north east, xshift=-2mm, yshift=-1.5mm,
            font=\scriptsize\upshape\color{black!70},
            inner sep=0pt] at (frame.north east) {#1};
    }
  },
  myprompt base/.style={
    fontupper=\upshape,
    boxrule=0.5pt,
    arc=2pt,
    left=2mm,
    right=2mm,
    top=3mm,
    bottom=2mm,
    breakable,
    enhanced,
    frame code={
      \path[draw=tcbcolframe, dashed, line width=0.6pt, rounded corners=2pt]
        (frame.south west) rectangle (frame.north east);
    },
    attach boxed title to top left={xshift=2mm,yshift*=-2mm},
    coltitle=black,
    boxed title style={
      size=small,
      colframe=black,
      boxrule=0.5pt,
      arc=2pt,
    }
  }
}

\newcommand{\defquotebox}[3]{%
  \newtcolorbox{#1quotebox}[1]{
    myquote base,
    colback=#2!5,
    colframe=#2!50!black,
    title={Quote: #3},
    boxed title style={colback=#2!50},
    model tag=##1
  }
}

\newcommand{\defpromptbox}[3]{%
  \newtcolorbox{#1promptbox}{
    myprompt base,
    colback=#2!5,
    colframe=#2!50!black,
    title={Prompt: #3},
    boxed title style={colback=#2!50}
  }
}

\defquotebox{as1}{ForestGreen}{AS1}
\defpromptbox{as1}{ForestGreen}{AS1}

\defquotebox{as2}{RoyalBlue}{AS2}
\defpromptbox{as2}{RoyalBlue}{AS2}

\defquotebox{user}{BurntOrange}{User}
\defpromptbox{user}{BurntOrange}{User}

\defquotebox{university}{Maroon}{University}
\defpromptbox{university}{Maroon}{University}

\defquotebox{acm}{Violet}{ACM}
\defpromptbox{acm}{Violet}{ACM}

\defquotebox{eveink}{Teal}{EveLink}
\defpromptbox{eveink}{Teal}{EveLink}

\defquotebox{transitas}{Goldenrod}{TransitAS}

\defquotebox{isp}{Brown}{ISP}
\defpromptbox{isp}{Brown}{ISP}

\lstdefinelanguage{rawtext}{}
\newtcbinputlisting{\promptbox}[2]{%
    enhanced,
    breakable,
    listing only,
    listing options={
        language=rawtext,
        basicstyle=\itshape\small\ttfamily,
        breaklines=true,
        breakatwhitespace=true,
        breakautoindent=false,
        columns=fullflexible,
        keepspaces=true,
        breakindent=0pt,
        upquote=true,
        extendedchars=true,
        literate=
          {—}{{-{}-{}-}}3
          {–}{{-{}-}}2
          {‘}{{`}}1 {’}{{'}}1 {“}{{"}}1 {”}{{"}}1
          {…}{{...}}3
    },
    colback=boxbg, colframe=boxframe, rounded corners, arc=5pt, boxrule=1pt,
    top=10pt, before skip=15pt, after skip=15pt,
    attach boxed title to top left={yshift=-6pt, xshift=10pt},
    boxed title style={colback=titlebg, colframe=boxframe, arc=3pt, boxrule=1pt,
        top=1pt, bottom=1pt, left=6pt, right=6pt},
    coltitle=white, fonttitle=\bfseries\sffamily,
    title={#1},
    listing file={#2}
}
 
\definecolor{reportbg}{HTML}{F7F4FA}
\definecolor{reportframe}{HTML}{8264A6}
\definecolor{reporttitlebg}{HTML}{9C7FBC}

\newtcolorbox{reportboxinner}{%
    enhanced, breakable,
    colback=reportbg, colframe=reportframe,
    rounded corners, arc=5pt, boxrule=1pt,
    top=10pt, before skip=15pt, after skip=15pt,
    attach boxed title to top left={yshift=-6pt, xshift=10pt},
    boxed title style={colback=reporttitlebg, colframe=reportframe, arc=3pt, boxrule=1pt,
        top=1pt, bottom=1pt, left=6pt, right=6pt},
    coltitle=white, fonttitle=\bfseries\sffamily, title={Report}
}

\newcommand{\paragraphspace}[0]{\vspace{2pt}}

\newcommand{\jpara}[1]{%
\paragraphspace%
  \noindent\textbf{{#1}.}
}

\newcommand{\eat}[1]{}

%% file: sections/abstract.tex
\begin{abstract}
In 2003, Clark et al. proposed the ``Knowledge Plane'', a layer of intelligent agents that share rich information about network state in order to collaboratively manage the Internet across administrative boundaries.
The technology of the day lacked the two capabilities it demanded most: \emph{reasoning} and \emph{unstructured communication}. Thus, automated network management has remained largely confined to single domains, while interdomain problems are still resolved by human operators on the phone.
We argue that agentic large language models can finally unlock the potential of the Knowledge Plane,
but that doing so will require solving new research problems that arise particularly in this decentralized, multi-agent, coopetitive setting that have received limited attention in the literature thus far.
\end{abstract}

%% file: sections/intro.tex
\section{Introduction}
The Internet is designed to be general, composable, and decentralized -- properties that enable it to scale globally. However, \emph{managing} components of the Internet can be a formidable task precisely because of its decentralized nature and complexity -- so much so that entire subfields have been devoted to network configuration, routing, troubleshooting, defense, load balancing, and more~\cite{benson2009unraveling,kim2013improving,beckett2017general,gao2001stable,mahajan2002understanding}. These challenges currently require a great deal of human coordination on top of the sophisticated algorithms, software, and hardware built to solve these problems. 

In the face of so much complexity and manual intervention, one would naturally ask: what can we hope to automate away? Over decades of effort, researchers have proposed numerous solutions to automate the configuration and verification of pieces of network management~\cite{feamster2005detecting,fogel2015general,beckett2017network,el2018netcomplete,tian2019safely,ramanathan2023practical,kazemian2012header}. Yet protocols for rich \emph{interdomain}
communication are limited, 
and difficult to upgrade and adapt for unforeseen applications --
 so automated network debugging and decision-making
tends to be limited to \emph{single-domain} solutions.

Presciently, in 2003, Clark et al. proposed a solution: a ``Knowledge Plane'' run by intelligent agents that share rich information about network state with the goal of collaborative network management, including troubleshooting, routing, configuration, intrusion detection and more~\cite{clark2003knowledge}. The paper envisioned a user bringing a question like, ``Why can't I load this webpage?'' to an agent which might in turn collect data about routes and DNS from its local network, and ask agents from remote networks about the location of their firewalls or their provisioned server capacities. In the pre-LLM era, however,  it was not clear \emph{how} one would implement a Knowledge Plane. Although several lines of subsequent work have attempted to instantiate a Knowledge Plane \cite{mestres2017knowledge,self-driving-net,ashtari2022knowledge,wijesekara2023comprehensive,bullot2008situatedness,macedo2008knowledge}, 
the state of the art technology at the time lacked sufficient \emph{unstructured communication} and \emph{reasoning} capability to autonomously and collaboratively diagnose and solve interdomain network problems. It is only now that we have models -- agentic large language models -- powerful enough to truly instantiate a Knowledge Plane.

The natural instinct of the skeptical systems researcher may be to push back against the idea of unleashing unpredictable agents on the Internet, or perhaps to question the technical value of an agentic layer over the Internet. We acknowledge these concerns, yet hope to argue that LLM agents can fill a real technical need that current Internet protocols and automated troubleshooting solutions do not adequately solve, while also outlining concrete challenges -- particularly in safety and security -- that must be met in order for deployment to be possible. In short, we ask the reader to temporarily ignore the question of whether we \textit{should} manage the Internet with agents, and join us for the thought experiment of what benefits and challenges we might uncover if we were to do so.

\jpara{Promise: Decentralized, online, adaptive decision making} 
 Many real problems require communication across network domains to troubleshoot. To name a few examples, diagnosing protocol failures due to unforeseen middlebox behavior~\cite{langley2017quic, raiciu2012mptcp}, diagnosing path inflation~\cite{spring2003pathinflation}, identifying and root-causing interdomain congestion~\cite{dhamdhere2018congestion}, and identifying unreachable ``black holes'' ~\cite{katzbassett2008hubble}
are all examples of distributed network failures that cannot be easily diagnosed nor resolved with the information available in a single autonomous system.

Cooperating LLM agents hosted by every network, following the Knowledge Plane's blueprint, bring two main benefits: decentralized collaboration and adaptive communication.

First, LLMs can \emph{collaborate} to solve complex troubleshooting tasks or make decisions involving multiple stakeholders. Today, such problems would require multiple human operators to coordinate, requiring potentially several communication rounds and hours of human effort to debug. Importantly, even though some automated tools can be used to inspect logs or perform data analysis and speed up this process to a degree, \emph{interdomain} network management is difficult because no individual operator has access to all of the data, infrastructure, and configurations that a route or packet is going to touch as it crosses the Internet.

Second, LLMs can \emph{adapt} their communication as needed by the application at hand: natural language functions as a flexible and completely forward- and backward-compatible protocol. Fixed protocols to exchange telemetry between ASes (e.g. netflow data, routing configuration) between operators might be helpful, but they ultimately constrain the data that is shared: new protocols or systems might not be able to be encoded in the inter-AS telemetry format as one AS adopts a new technology and another lags behind. An agentic model brings the ability of a sender to {\it explain} what new data represents, even as new technologies are adopted. Furthermore, operators may have privacy concerns and hence not {\it want} to share arbitrary telemetry data unless the data is explicitly necessary to solve a problem at hand. Deciding which data to share requires reasoning, something that LLMs might resolve but a static protocol with traditional ACLs cannot. 

In \cref{sec:case-studies}, we present a series of simulated case studies where agents at different organizational domains collaborate to debug availability failures, defend against security risks, and optimize financial goals. This illustrates the capacity of today's agents to communicate, collaborate, and resolve the kinds of interdomain issues that today force human network operators to be available on the phone around-the-clock.

\jpara{Challenge: Collaboration without trust} In \cref{sec:discussion}, we explore why an agentic Knowledge Plane is not (yet?) a good idea. Some open challenges are shared with many other systems that hope to adopt LLM technology for better management, especially challenges in scalability~\cite{chen2023frugalgpt, kolawole2024agreement, ong2024routellm, jitkrittum2023confidence, soiffer2025semantic, kossmann2024cascadeserve, taalas2026company} and LLM correctness and robustness to attacks~\cite{staufer20262025, greenblatt2024aicontrol, baker2025monitoring, benton2024sabotage, kutasov2025shadearena, inan2023llamaguard, kang2024r2guard, rebedea2023nemo, xiang2024guardagent}. 

At the same time, we think an agentic Internet brings challenges that are uncommon to the broader research world, as we expect agents to {\it collaborate} towards a shared goal (global routing) despite a lack of {\it trust} between collaborating agents as well as potentially conflicting \emph{local} goals from diverse edge applications. This plays out in two open questions we see: data privacy (operators do not generally share all of their routing preferences, and yet agents may need to share some of  this information to diagnose issues) and strategic behavior (untrusted networks may selectively withold information or even lie). As we will discuss further in \cref{sec:discussion}, these kinds of challenges are less well-explored in either the ML literature or the literature of applying ML to computer systems. Hence, an agentic Internet brings new research perspectives into AI safety and multiagent coordination.

%% file: sections/defining-kp.tex
\section{Background and Related Work}
\label{sec:defining-kp}

In this section, we present the original vision of the
Knowledge Plane (KP) as envisioned by Clark et al. ~\cite{clark2003knowledge},
review significant literature that has made strides towards realizing that vision since 2003,
highlight the remaining gaps to realizing that full vision,
and outline concrete requirements that our agentic architecture should satisfy in order to realize the interdomain vision of the Knowledge Plane.

\subsection{The Knowledge Plane Vision}
The goal of the Knowledge Plane was to create ``a
fundamentally different sort of network that can assemble itself
given high level instructions, reassemble itself as requirements
change, automatically discover when something goes wrong, and
automatically fix a detected problem or explain why it cannot do so''~\cite{clark2003knowledge}.
In the paper, the authors argue that a ``cognitive system'' is necessary to bring {\it reasoning} into the management of Internet infrastructure (despite the limitations of AI reasoning systems at the time). At the heart of the paper is an example -- which we replicate in \cref{sec:case-studies} -- in which a user asks a broad and high level question (``Why can't I access acm.org?'') and a distributed collection of agents from the user's AS, the ACM's AS, and even agents at individual servers exchange information about routes, DNS availability, firewalls, server capacity, etc.

Today, debugging an outage with this level of data availability would require phone calls between operators. However, the phone-call approach is only feasible for major CDNs, networks, and datacenter operators to use in only the worst-case of outages -- and hence most individual problems of this nature are never resolved at all. Hence, the KP approach represented then, and still represents today, a major advance in the ability to diagnose and potentially repair interdomain challenges.

\subsection{Then and Now}

Some have attempted to explicitly instantiate a 
Knowledge Plane~\cite{mestres2017knowledge,self-driving-net,ashtari2022knowledge,wijesekara2023comprehensive,bullot2008situatedness,macedo2008knowledge},
but prior to LLMs,
the required capabilities for reasoning and unstructured communication were not available and hence these approaches lacked the ``cognitive'' depth envisioned by Clark et al.

Other lines of work in the intra-domain setting have pushed the reasoning aspects of a Knowledge Plane approach further.
For example, work on \emph{self-driving networks}~\cite{self-driving-net, gupta2018sonata, saeed2019eiffel, usama2019unsupervised, kfoury2021exhaustive, swamy2022taurus, akem2023flowrest, sivaraman2016programmable}
and related work on machine learning for network management applications, particularly reinforcement learning~\cite{mao2016resource, mao2019learning, mao2017neural}, are particularly rich.
The seminal paper on self-driving networks~\cite{self-driving-net} outlined a number of goals of
network automation, covering data collection,
in-network analytics and decision making,
security, scalability, learning, and dynamic decision making.
We see many of these contributions as
\emph{supportive} of the proposal of an agentic Knowledge Plane: these works illustrate a division of labor in which
agents can reason and communicate,
but the job of data collection and analytics are still best left
to the programmable data plane and traditional statistics,
which agents can access as tools. 

Significant progress has been made towards leveraging LLMs in the context of network management.
One line of work proposes \emph{intent-based networking} \cite{clemm2022intent, avgerinos2025arm, wang2025confucius, mekrache2024intentcentric, wang2025confucius, ramanathan2023practical, tian2019safely, jacobs2018refining, jacobs2021hey} in which network operators should be able to express \emph{goals and outcomes} (in some interface such as, but not limited to, natural language) and the network should translate and accomplish that intent -- but the goal is typically to translate user intent into network behavior; network devices cannot necessarily send unstructured expressions to each other.
Agentic techniques have already begun to be incorporated into network and software system troubleshooting \cite{aykurt2024netllmbench, roy2024exploring, sade2026, tnautorca2025, networkarena2025, wang2025bian, wang2024netassistant, wang2024rcagent, wang2025confucius, xu2025openrca, bothra2024lightweight}.

Altogether, the Knowledge Plane vision is far more feasible today than it was in 2003, in part due to advances in LLM technology, but also due to advances in the network management literature as described above.

\subsection{Requirements for a Knowledge Plane}

We now consider the core properties that a Knowledge Plane requires and what remains unsolved for the interdomain Knowledge Plane vision to be realized.
From Clark et al. ~\cite{clark2003knowledge},
we derive the following list of requirements:

\begin{enumerate}[label=\textbf{R\arabic*}]
    \item \textbf{Abstract high-level goals from low-level actions.}

    \item \textbf{Learn from past actions.}

    \item \textbf{Operate effectively in the face of generality.}

    \item \textbf{Extend its perspective to the entire global network by merging perspectives and activities between different networks.}

    \item \textbf{Function in the presence of incomplete, inconsistent, misleading, or malicious information.}

    \item \textbf{Perform appropriately in the presence of conflicting or inconsistent higher-level goals among the Internet's different stakeholders.}
\end{enumerate}

Among these requirements,
we observe that R1-R3 largely test
\emph{single-agent} capabilities that can be explored within a single domain.
{\it The existing literature on AI-based network management largely solves all of them!}
Taking one recent and prominent work as an example, Meta's recent paper~\cite{wang2025confucius} demonstrates a {\it production} tool that moves between low level data formats and higher level human abstractions (R1), learns and models anomalies (R2), and adapts as new tools and technologies are deployed (R3).

On the other hand, the remaining R4-R6 test \emph{multi-agent} interactions, which is required for multi-domain settings. 
These capabilities have not yet been demonstrated in the literature. 

To explore this possibility, in the following section,
we select three case studies chosen to illustrate that today's LLMs are capable of at least some of the reasoning and communication 
requirements detailed in R4-R6, pointing the way towards the possibility of an interdomain Knowledge Plane.

%% file: sections/case-studies.tex
\section{Case Studies}
\label{sec:case-studies}

In this section, we walk through a series of case studies to better understand whether agents would meet the three criteria we chose to study in \cref{sec:defining-kp}.
We first (\cref{sec:fault-diagnosis}) simulate an example from the KP paper~\cite{clark2003knowledge}, testing the agents' ability to reason about a global perspective (R4).
We then (\cref{sec:bgp-example}) illustrate a scenario in which the LLMs demonstrate capabilities to consider conflicting and even malicious information (R5 and R6).
Finally, we consider a more dynamic setting in which agents can perform unconstrained load balancing (\cref{sec:agentic-routing-policies}).

\subsection{Experimental Setup}

We use Mininet~\cite{mininet} to build a virtualized network of hosts, each equipped with an agent acting as a system administrator. Agents treat the emulated network as a real one, issuing real commands on their respective hosts in the Mininet environment. Experiments run in simulated time, so scenarios spanning days of network activity complete in minutes.

Each agent has full administrative control over its own host, and is informed of its immediate neighbors, but has no global view of the network. Each agent receives a prompt that briefly describes its organization's objectives and policies. We keep these prompts minimal: enough to act on the organization's behalf, but without details of the experimental scenario. In particular, agents are never told which fault was injected, or that one was injected in the first place. Agents communicate outside the data plane, over a separate dedicated message channel (the Knowledge Plane), which they use to send messages to their immediate neighbors.

Each agent is provided with the same set of tools: \texttt{exec} to run a shell command on their host, \texttt{send\_message} to communicate in natural language with other neighboring agents via the Knowledge Plane, and \texttt{idle} to wait until a message arrives. Every tool usage carries a mandatory free-text justification, which is the source of the reasoning quoted throughout this section. An experiment ends once every agent signals it has nothing left to do, or when a fixed budget of iterations is exhausted.

We ran all experiments with four models: Claude Opus 4.7, GPT 5.5, and smaller open-weight models Qwen 2.5 72B (AWQ) (with vLLM~\cite{kwon_efficient_2023} and two NVIDIA RTX PRO 6000 Blackwell Max-Q) and GLM 5.2 FP4 (via Together AI~\footnote{\url{https://www.together.ai}}).

All our code and experimental results, including prompts, logs, and reports, are openly available~\footnote{ \url{https://fchamicapereira.github.io/agentic-networks/}}.

\begin{figure}[t]
\includegraphics[width=1\linewidth]{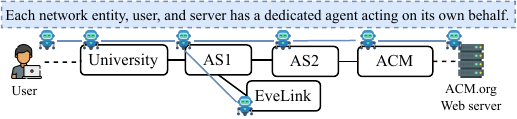}
\caption{Simulated network for fault diagnosis wherein a student tries to access \texttt{www.acm.org}.}
\label{fig:knowledge_plane}
\end{figure}

\subsection{Fault Diagnosis}
\label{sec:fault-diagnosis}

Consider an example proposed by Clark et al. ~\cite{clark2003knowledge}, wherein a university student cannot reach \texttt{www.acm.org} (\cref{fig:knowledge_plane}). The cause may lie anywhere along the path: a firewall block at the university, a routing failure upstream, a breakdown in name resolution, or even a server overload at ACM. Diagnosing it requires \emph{cooperation across networks} with different vantage points but under \emph{separate administrative control}. 

We investigate our Knowledge Plane's ability to recognize and recover from an outage of \texttt{www.acm.org} under three different causes. We initialize each agent with the basic rules and responsibilities of its organization.
For instance, for AS1:

\begin{as1promptbox}
You belong to the Knowledge Plane, and represent autonomous system AS1. University and EveLink are your customers, and pay you for Internet transit. AS2 is your peer, and neither one of you pays the other (you exchange traffic as equals). Your goal is to maximize revenue, and provide reliable Internet transit to your customers.
\end{as1promptbox}

The student always takes the first step by asking their local agent for help. The agent then collaborates with the rest of the Knowledge Plane to finally return a diagnosis.

\jpara{Cause \#1: University firewall} The university administration team mistakenly introduced a firewall rule that inadvertently blocked requests to \texttt{www.acm.org}. For all models, the user agent first replicated the reported issue, ran local diagnostics, and \emph{escalated the problem} automatically to the university. The university agent then found the smoking gun, which was later relayed back to the user:

\begin{universityquotebox}{GLM 5.2 FP4}
Firewall policy on the University gateway (Uni) is blocking all traffic to 198.82.0.0/24, which includes 198.82.0.1 (acm.org). [...] Per policy, changes to firewall rules require administrator approval — I cannot override them autonomously. Administrators have been notified of this finding and the proposed fix.
\end{universityquotebox}

\jpara{Cause \#2: AS1's stale DNS entry} Now, the user's requests pass freely through the university's firewall, but AS1 is configured with an out-of-date DNS entry resolving \texttt{www.acm.org} to an old ACM IP address. As the user's machine relies on AS1 to resolve \texttt{www.acm.org}, every subsequent HTTP request to that website is forwarded to the wrong location.

Opus, GPT, and GLM quickly escalated the problem up to AS1, which correctly identified the DNS misconfiguration.
However, it deemed the required change sensitive enough to require administration approval, notifying the administrators and warning University of the findings, which ultimately relayed to the user:

\begin{userquotebox}{Claude Opus 4.7}
acm.org failed to load because our upstream DNS provider is handing out a wrong address for the site [...]. The issue has already been flagged to that provider's administrators, so no action is needed from you on that front.
\end{userquotebox}

Qwen, on the other hand, gathered the same evidence but read it as a routing issue rather than a DNS one. It was ultimately unable to provide an accurate report to the user.

\jpara{Cause \#3: Overloaded ACM server} In this scenario, neither the university's firewall nor AS1 are the source of the outage. We instead exhaust the ACM web server's capacity by having EveLink issue \emph{benign} requests to a blocking HTTP endpoint of the ACM webserver. The server remains reachable, but rejects new connections.

Both Opus and GPT escalated to Web and correctly identified overload as the root cause. Web then shared its findings with ACM, including client information and implementation details. ACM relayed the diagnosis back to the user but, as instructed, withheld the private implementation details and the fact that EveLink was the reason behind this outage.

\begin{acmquotebox}{Claude Opus 4.7}
I am NOT disclosing the abusive source IP, the /slow endpoint, worker-pool details, or any other internal specifics to external parties — only that the service is degraded, cause is identified as external abuse, and mitigation is pending admin approval.
\end{acmquotebox}

Qwen struggled to correctly diagnose the problem, not escalating to ACM and thus never collecting enough information to reach a verdict and inform the user.
GLM instead read EveLink's behavior as malicious (``slow-loris-like attack'').
But because ACM's directives forbade patching its firewall directly, GLM exposed EveLink's actions publicly and asked AS1 for help, overriding its privacy directives to do so.
AS1 relayed this to EveLink, which stopped, but EveLink was still blamed for the incident in the user's report.

\jpara{Key takeaway} In all three experiments, the agents cooperated across organizational boundaries, reasoning together and merging partial views of the network into a single diagnosis, providing evidence that they can fulfill R4.

\subsection{Reasoning amidst conflicting information}
\label{sec:bgp-example}

\begin{figure}[t]
\includegraphics[width=.8\linewidth]{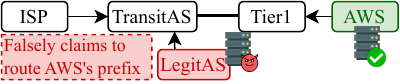}
\setlength{\belowcaptionskip}{-2em}
\caption{Topology of the Celer Bridge experiment. Arrows denote provider/client relationship, and solid lines peer/peer.}
\label{fig:celer_bridge}
\end{figure}

We now turn from Clark et al.'s examples to an adversarial scenario, testing how agents handle conflicting information and high-level goals (R5, R6) by simulating the 2022 Celer Bridge BGP hijack~\cite{coinbase2022celer,madory_what_2022}. \Cref{fig:celer_bridge} shows the topology of our simulation, analogous to the Celer Bridge incident.

In this incident, the attackers served a legitimate-looking but malicious front-end to users of Celer Bridge, a service for moving cryptocurrency assets between blockchains.
The attack hinged on a BGP hijack: a small hosting provider (LegitAS) announced it could reach a block of AWS addresses hosting the legitimate front-end, drawing traffic to attacker-controlled servers.
The announcement was crafted to pass the routing system's origin checks (IRR~\footnote{The Internet Routing Registry (IRR)~\cite{mcpherson_considerations_2015,noauthor_representation_1995} is a set of databases that associate ASs with the IP space they are authorized to originate.} and RPKI~\footnote{The Resource Public Key Infrastructure (RPKI) cryptographically authorizes ASs to originate prefixes~\cite{lepinski_rfc_nodate}.}), so no protection flagged it.

Qwen failed to stop the attack, accepting RPKI-validity as enough evidence to install the malicious announcement. All of the other models rejected LegitAS's announcement on three pieces of evidence: first, the fact that some IRR entries were not verified; second, the implausible implication that a small AS such as LegitAS has a direct connection with AWS and is authorized to transit traffic to them; third, AWS's agent denied authorizing the IRR entry LegitAS relied on, exposing it as fabricated rather than merely unverified.

\begin{transitasquotebox}{GPT 5.5}
I am not installing or exporting this more-specific at this time. The prefix is within AWS space already reachable via a validated 44.192.0.0/16 route, and the supporting IRR objects/as-set are AltDB-only; RPKI validates only the origin/max-length, not that AS-LEGITAS is authorized to provide transit for AS-AWS. [...] I am keeping 44.192.0.0/16 via Tier1 and requesting stronger AWS authorization before considering acceptance.
\end{transitasquotebox}

\jpara{Key takeaway} Skepticism in the face of incorrect advertisements provides evidence that agents can reason about misleading information (R5) and conflicting information and goals (R6).

\subsection{Agentic policy-based routing}
\label{sec:agentic-routing-policies}

We now move beyond fault analysis, and ask whether agents can also make \emph{dynamic decisions} in a load-balancing setting.

In particular, instead of relying on BGP, we allow agents to autonomously make routing decisions.
Crucially, each agent is provided with the routing contracts signed off by its organization.
Unlike BGP configurations, which strive for mimicking details of contractual relationships (expressed in natural language by providers) in a programmatic language, the agents now have full access to the nuances of each contract.

We analyze the routing decisions made by an ISP connected to two providers: Cheap and Expensive. While Cheap charges \$1/Mbps, Expensive provides guaranteed low-latency transit at \$5/Mbps, billed at the $90^{\text{th}}$ percentile over a 30-day billing period. ISP is instructed to minimize cost.

\begin{figure}[t]
\includegraphics[width=0.8\columnwidth]{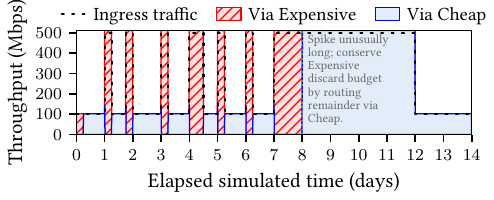}
\caption{ISP routing decisions between Cheap and Expensive when running with Opus or GPT. The agent reasons about the $90^{\text{th}}$ percentile billing on the Expensive option, rerouting traffic spikes to it without getting billed.}
\label{fig:billing}
\end{figure}

\Cref{fig:billing} shows 14 days of traffic flowing through ISP when running with either Opus or GPT.
Although during common 100Mbps windows traffic is routed through Cheap, the few hours-long daily spikes of traffic are routed through Expensive.
The agent's reasoning is founded on the fact that the busiest 10\% of samples are discarded from Expensive's billing, allowing it to exploit Expensive's superior service quality to absorb these temporary spikes without paying the additional cost.
After day 7, the traffic diverges from the pattern, perpetuating the high volume for several days.
Within a day or two, the agent reverts to Cheap, as sustaining the elevated traffic through Expensive would exhaust its limited discard budget.

\begin{ispquotebox}{Claude Opus 4.7}
This is the 5th consecutive spike sample. Very long spike. I need to think again about whether my budget management is on track. [...] Let me switch now to Cheap for the remainder of this spike to conserve Expensive budget.
\end{ispquotebox}

Qwen showed no evidence of reasoning about percentile billing, arbitrarily swapping between Cheap and Expensive.
GLM reasoned about the percentile billing at the start, rerouting only traffic spikes through Expensive. However, once the Cheap top-percentile budget was depleted it wrongly decided to permanently route through Cheap, as ``routing through Expensive would only increase the bill''.

\jpara{Key takeaway} The reasoning power of the models allows them to take actions that directly reflect the contracts they are provided with. This not only strongly aligns with the Knowledge Plane vision, wherein agents make low-level decisions well aligned with the high-level goals, but lets us peek into a future where agents actively make decisions in the network, beyond providing diagnostic support.

%% file: sections/discussion.tex
\section{Discussion and Challenges}
\label{sec:discussion}

We presented a proof of concept that
demonstrates that a Knowledge Plane is both \emph{feasible} using
modern LLMs and \emph{useful} in the sense that an autonomous, language-based reasoning layer can adapt and react to unforeseen circumstances in a way that fixed protocols cannot, and can do so without expensive protocol updates that require wide consensus to deploy.

However, given that LLMs are still a relatively nascent and unpredictable technology, such a design raises many obvious concerns.
Some of these are general concerns that are actively being investigated by several research communities and apply to many downstream applications of LLM agents.
Other concerns are more specific to the Internet Knowledge Plane use-case.
In general, we see these as rich areas for debate as well as further research.

\subsection{Challenges Shared by Other Agentic Systems}

\jpara{Resource cost}
Our prototype assumed that each entity in the network had access to a large frontier-capability model
hosted remotely and accessed via an API -- an architecture that would be prohibitively expensive (in cost, compute, and latency) at Internet scale.

Solving this problem would likely involve a combination of approaches.
We can deploy agents at a coarser granularity (e.g., one per AS) to manage a collection of devices.
We can also vary the power and cost of the model using \emph{model cascades}, which dynamically select the required model size based on properties of the queries at hand and cost constraints \cite{chen2023frugalgpt, kolawole2024agreement, ong2024routellm, jitkrittum2023confidence, soiffer2025semantic, kossmann2024cascadeserve}.
\emph{Open models} (such as GLM) give significantly more control over model serving, latency, and data sharing compared to closed models, and can perform comparably to closed models~\cite{gaba2026benchmarking}.
An additional, promising direction is that of custom silicon for ultra-fast serving of open-weight models  \cite{taalas2026company}. This custom hardware can enable low-latency, low-cost, and local serving of models on-device.

In a decentralized network,
organizations would be able to \emph{choose}
which models to deploy on their devices.
This leverages one of the main strengths of agentic communication: while agents may run on different hardware or be trained differently, their ``protocol'' (i.e., natural language) will be compatible.
Nevertheless, heterogeneous agents would likely have to provide sufficient detail about their reasoning process to convince neighboring agents that their decision making is trustworthy.

\jpara{Constraining agent behavior}
Two significant challenges in deploying any agent are (1) ensuring the predictability and safety of the agent's actions and (2) preventing attacks aimed at modifying an agent's behavior in malicious ways. 
These problems are far from solved, but they are shared concerns with almost all applications of LLM agents, and are under active study and research in the machine learning community.

A common mitigation is to introduce an agentic monitor~\cite{staufer20262025, greenblatt2024aicontrol, baker2025monitoring, benton2024sabotage, kutasov2025shadearena, bhatt2025ctrlz, pacchiardi2023liar, kenton2024scalable, bowman2022measuring} (a watchdog agent to evaluate behavior), and increasing work examines \emph{scalable oversight}, in which a smaller model judges a larger model, making monitoring practical~\cite{kenton2024scalable, bowman2022measuring, kale2025reliable, chan2025infrastructure}.
Other common solutions include guardrails that filter harmful output \cite{inan2023llamaguard, kang2024r2guard, rebedea2023nemo, xiang2024guardagent} and, increasingly, enforcing access control via safety policies~\cite{omnigent_policies_overview}.
A number of other works look at protecting agents from adversarial attacks including prompt injection~\cite{debenedetti2025camel, wu2024fsecure, debenedetti2024agentdojo} and jailbreaking~\cite{benton2024sabotage, kutasov2025shadearena}.

While the above safeguards naturally carry over to network management, the networking setting also introduces new challenges -- e.g., sandboxing or limiting the impact of a decision in a distributed system, limiting private data sharing in a situation that naturally requires information sharing, preventing attacks that are unique risks in a distributed Internet domain (particularly prompt injection-style attacks).
On the other hand, this domain also has unique opportunities to restrict behavior that may not be available in other domains -- an agent running in a router does not need the full capability of, for example, a coding agent developing networked software.

\subsection{Challenges Distinctive to the Interdomain Setting}

\jpara{Adaptive and selective information sharing}
Interdomain management fundamentally involves sharing data and information.
How should independently operated agents decide what information to share, with whom, and under what circumstances?

Answering this question requires both reasoning about the problem
at hand and \emph{what information} is minimally required to solve a problem,
and also reasoning about the \emph{privacy} risks of sharing user data
among different domains. 
A benefit of structured protocols is that all parties can agree on a schema for data and information to be shared in advance --
but the downside is that these protocols can be limiting and can ossify quickly. 
Unstructured agentic communication can overcome these limitations,
but agents must then reason about sophisticated trust relationships and
privacy considerations, not all of which may be made explicit by end-users.
Some progress has been made on mitigations for this problem in the context of agents within an organization that have access to many users' data~\cite{omnigent_policies_overview, wiselman2026memorysafety},
but tracking data provenance and managing access at an Internet scale is a new challenge. 

\jpara{Cooperation among mutually distrusting agents}
A number of existing works in the agentic LLM literature look
at using ensembles of agents that cooperate to solve parts of a single overarching task \cite{wang2025confucius, bhattacharyya2026social, chen2025harnessing, agashe2025llm}.
The Internet setting is different:
decentralized agents that do not have full trust in each other
attempt to cooperate, but to achieve individual objectives,
which may not be known to each other, and may not even be consistent.
Even in other scenarios composed of decentralized entities with
their own agents, such as robot swarms,
typically these work together for a collective goal rather than potentially conflicting goals.

As a result, the Internet domain truly presents a new problem statement that is largely unexplored in the machine learning literature:
decentralized negotiation among agents. 
Progress on this problem would likely be of interest not only to the
networking community but also result in general improvements in agent capabilities applicable to other domains (such as coordination of autonomous vehicles or economic modeling). 
Notably, moving forward on this problem would first require identifying
suitable metrics to measure success and developing benchmarks to evaluate progress and structure and unify research.

Moreover, effective agents in this domain 
will need to be able to reason about complex 
problems potentially involving a multitude of devices and domains. 
A single agent may or may not be able to effectively reason about problems at an Internet scale.
Solutions to this problem may take many forms:
more capable individual agents, escalation to more-powerful compute backends when necessary,
or even efficient information-sharing systems.
The good news is that the Internet was built on an ethos of decentralized reasoning, so researchers and practitioners have decades of intuition on how to effectively make use of limited and decentralized communication, carrying these lessons over into an agentic knowledge plane.

%% file: sections/conclusion.tex
\section{Conclusion}

As early as 2003, Clark et al. predicted that a cross-domain reasoning and communication layer would be key to lifting the heavy burden of network management from human operators.
LLMs represent a step-change in the reasoning and communication capabilities of our computing devices, and make this vision possible. 
While significant and important challenges remain in achieving this vision in practice, 
those challenges are worthy agendas and solving them would advance both networking and machine learning research.

%% file: sections/acknowledgments.tex
\section*{Acknowledgments}

We are grateful to our shepherd, Gábor Rétvári, and the anonymous HotNets'26 reviewers. We thank Hugo Sadok and Matei Zaharia for comments on earlier drafts of the paper.
Justine Sherry holds concurrent appointments as
an Associate Professor at CMU and as an Amazon Scholar.
This work was conducted at CMU and is not affiliated with Amazon.
This work was supported by a Sloan Research Fellowship (FG-2024-21263) and by the ONR under grant \#N00014-24-1-2059.
Francisco Pereira was supported by the FCT scholarship PRT/\allowbreak BD/\allowbreak 152195/\allowbreak 2021.

%% file: hotnets26.bbl

\begin{thebibliography}{88}


\ifx \showCODEN    \undefined \def \showCODEN     #1{\unskip}     \fi
\ifx \showISBNx    \undefined \def \showISBNx     #1{\unskip}     \fi
\ifx \showISBNxiii \undefined \def \showISBNxiii  #1{\unskip}     \fi
\ifx \showISSN     \undefined \def \showISSN      #1{\unskip}     \fi
\ifx \showLCCN     \undefined \def \showLCCN      #1{\unskip}     \fi
\ifx \shownote     \undefined \def \shownote      #1{#1}          \fi
\ifx \showarticletitle \undefined \def \showarticletitle #1{#1}   \fi
\ifx \showURL      \undefined \def \showURL       {\relax}        \fi
\providecommand\bibfield[2]{#2}
\providecommand\bibinfo[2]{#2}
\providecommand\natexlab[1]{#1}
\providecommand\showeprint[2][]{arXiv:#2}

\bibitem[Agashe et~al\mbox{.}(2025)]%
        {agashe2025llm}
\bibfield{author}{\bibinfo{person}{Saaket Agashe}, \bibinfo{person}{Yue Fan},
  \bibinfo{person}{Anthony Reyna}, {and} \bibinfo{person}{Xin~Eric Wang}.}
  \bibinfo{year}{2025}\natexlab{}.
\newblock \showarticletitle{Llm-coordination: evaluating and analyzing
  multi-agent coordination abilities in large language models}. In
  \bibinfo{booktitle}{\emph{Findings of the Association for Computational
  Linguistics: NAACL 2025}}. \bibinfo{pages}{8038--8057}.
\newblock


\bibitem[Akem et~al\mbox{.}(2023)]%
        {akem2023flowrest}
\bibfield{author}{\bibinfo{person}{Aristide Tanyi-Jong Akem},
  \bibinfo{person}{Michele Gucciardo}, {and} \bibinfo{person}{Marco Fiore}.}
  \bibinfo{year}{2023}\natexlab{}.
\newblock \showarticletitle{Flowrest: Practical flow-level inference in
  programmable switches with random forests}. In \bibinfo{booktitle}{\emph{IEEE
  INFOCOM 2023-IEEE Conference on Computer Communications}}. IEEE,
  \bibinfo{pages}{1--10}.
\newblock


\bibitem[Ashtari et~al\mbox{.}(2022)]%
        {ashtari2022knowledge}
\bibfield{author}{\bibinfo{person}{Sepehr Ashtari}, \bibinfo{person}{Ian Zhou},
  \bibinfo{person}{Mehran Abolhasan}, \bibinfo{person}{Negin Shariati},
  \bibinfo{person}{Justin Lipman}, {and} \bibinfo{person}{Wei Ni}.}
  \bibinfo{year}{2022}\natexlab{}.
\newblock \showarticletitle{Knowledge-defined networking: Applications,
  challenges and future work}.
\newblock \bibinfo{journal}{\emph{Array}}  \bibinfo{volume}{14}
  (\bibinfo{year}{2022}), \bibinfo{pages}{100136}.
\newblock


\bibitem[Avgerinos et~al\mbox{.}(2025)]%
        {avgerinos2025arm}
\bibfield{author}{\bibinfo{person}{Vasilis Avgerinos}, \bibinfo{person}{Kostas
  Ramantas}, \bibinfo{person}{Luis Alonso}, {and} \bibinfo{person}{Christos
  Verikoukis}.} \bibinfo{year}{2025}\natexlab{}.
\newblock \showarticletitle{{ARM}: Autonomous Remediation \& Management with
  {LLM} Agents for Intent-Driven Control}.
\newblock \bibinfo{journal}{\emph{IEEE Internet of Things Journal}}
  (\bibinfo{year}{2025}).
\newblock
\href{https://doi.org/10.1109/JIOT.2025.3648858}{doi:\nolinkurl{10.1109/JIOT.2025.3648858}}


\bibitem[Aykurt et~al\mbox{.}(2024)]%
        {aykurt2024netllmbench}
\bibfield{author}{\bibinfo{person}{Kaan Aykurt}, \bibinfo{person}{Andreas
  Blenk}, {and} \bibinfo{person}{Wolfgang Kellerer}.}
  \bibinfo{year}{2024}\natexlab{}.
\newblock \showarticletitle{Netllmbench: A benchmark framework for large
  language models in network configuration tasks}. In
  \bibinfo{booktitle}{\emph{2024 IEEE Conference on Network Function
  Virtualization and Software Defined Networks (NFV-SDN)}}. IEEE,
  \bibinfo{pages}{1--6}.
\newblock
\href{https://doi.org/10.1109/NFV-SDN61811.2024.10807499}{doi:\nolinkurl{10.1109/NFV-SDN61811.2024.10807499}}


\bibitem[Baker et~al\mbox{.}(2025)]%
        {baker2025monitoring}
\bibfield{author}{\bibinfo{person}{Bowen Baker}, \bibinfo{person}{Joost
  Huizinga}, \bibinfo{person}{Leo Gao}, \bibinfo{person}{Zehao Dou},
  \bibinfo{person}{Melody~Y. Guan}, \bibinfo{person}{Aleksander Madry},
  \bibinfo{person}{Wojciech Zaremba}, \bibinfo{person}{Jakub Pachocki}, {and}
  \bibinfo{person}{David Farhi}.} \bibinfo{year}{2025}\natexlab{}.
\newblock \showarticletitle{Monitoring reasoning models for misbehavior and the
  risks of promoting obfuscation}.
\newblock \bibinfo{journal}{\emph{arXiv preprint arXiv:2503.11926}}
  (\bibinfo{year}{2025}).
\newblock


\bibitem[Bates et~al\mbox{.}(1995)]%
        {noauthor_representation_1995}
\bibfield{author}{\bibinfo{person}{T. Bates}, \bibinfo{person}{E. Gerich},
  \bibinfo{person}{L. Joncheray}, \bibinfo{person}{J-M. Jouanigot},
  \bibinfo{person}{D. Karrenberg}, \bibinfo{person}{M. Terpstra}, {and}
  \bibinfo{person}{J. Yu}.} \bibinfo{year}{1995}\natexlab{}.
\newblock \bibinfo{booktitle}{\emph{Representation of {IP} {Routing} {Policies}
  in a {Routing} {Registry} (ripe-81++)}}.
\newblock \bibinfo{type}{Request for {Comments}} RFC 1786.
  \bibinfo{institution}{Internet Engineering Task Force}.
\newblock
\href{https://doi.org/10.17487/RFC1786}{doi:\nolinkurl{10.17487/RFC1786}}


\bibitem[Beckett et~al\mbox{.}(2017a)]%
        {beckett2017general}
\bibfield{author}{\bibinfo{person}{Ryan Beckett}, \bibinfo{person}{Aarti
  Gupta}, \bibinfo{person}{Ratul Mahajan}, {and} \bibinfo{person}{David
  Walker}.} \bibinfo{year}{2017}\natexlab{a}.
\newblock \showarticletitle{A general approach to network configuration
  verification}. In \bibinfo{booktitle}{\emph{Proceedings of the Conference of
  the ACM Special Interest Group on Data Communication}}.
  \bibinfo{pages}{155--168}.
\newblock


\bibitem[Beckett et~al\mbox{.}(2017b)]%
        {beckett2017network}
\bibfield{author}{\bibinfo{person}{Ryan Beckett}, \bibinfo{person}{Ratul
  Mahajan}, \bibinfo{person}{Todd Millstein}, \bibinfo{person}{Jitendra
  Padhye}, {and} \bibinfo{person}{David Walker}.}
  \bibinfo{year}{2017}\natexlab{b}.
\newblock \showarticletitle{Network configuration synthesis with abstract
  topologies}. In \bibinfo{booktitle}{\emph{Proceedings of the 38th ACM SIGPLAN
  conference on programming language design and implementation}}.
  \bibinfo{pages}{437--451}.
\newblock


\bibitem[Benson et~al\mbox{.}(2009)]%
        {benson2009unraveling}
\bibfield{author}{\bibinfo{person}{Theophilus Benson}, \bibinfo{person}{Aditya
  Akella}, {and} \bibinfo{person}{David~A Maltz}.}
  \bibinfo{year}{2009}\natexlab{}.
\newblock \showarticletitle{Unraveling the complexity of network management.}.
  In \bibinfo{booktitle}{\emph{NSDI}}. \bibinfo{pages}{335--348}.
\newblock


\bibitem[Benton et~al\mbox{.}(2024)]%
        {benton2024sabotage}
\bibfield{author}{\bibinfo{person}{Joe Benton}, \bibinfo{person}{Misha Wagner},
  \bibinfo{person}{Eric Christiansen}, \bibinfo{person}{Cem Anil},
  \bibinfo{person}{Ethan Perez}, \bibinfo{person}{Jai Srivastav},
  \bibinfo{person}{Esin Durmus}, \bibinfo{person}{Deep Ganguli},
  \bibinfo{person}{Shauna Kravec}, \bibinfo{person}{Buck Shlegeris},
  \bibinfo{person}{Jared Kaplan}, \bibinfo{person}{Holden Karnofsky},
  \bibinfo{person}{Evan Hubinger}, \bibinfo{person}{Roger Grosse},
  \bibinfo{person}{Samuel~R. Bowman}, {and} \bibinfo{person}{David Duvenaud}.}
  \bibinfo{year}{2024}\natexlab{}.
\newblock \showarticletitle{Sabotage Evaluations for Frontier Models}.
\newblock \bibinfo{journal}{\emph{arXiv preprint arXiv:2410.21514}}
  (\bibinfo{year}{2024}).
\newblock
\urldef\tempurl%
\url{http://arxiv.org/abs/2410.21514}
\showURL{%
\tempurl}


\bibitem[Bhatt et~al\mbox{.}(2025)]%
        {bhatt2025ctrlz}
\bibfield{author}{\bibinfo{person}{Aryan Bhatt}, \bibinfo{person}{Cody
  Rushing}, \bibinfo{person}{Adam Kaufman}, \bibinfo{person}{Tyler Tracy},
  \bibinfo{person}{Vasil Georgiev}, \bibinfo{person}{David Matolcsi},
  \bibinfo{person}{Akbir Khan}, {and} \bibinfo{person}{Buck Shlegeris}.}
  \bibinfo{year}{2025}\natexlab{}.
\newblock \showarticletitle{Ctrl-Z: Controlling AI agents via resampling}.
\newblock \bibinfo{journal}{\emph{arXiv preprint arXiv:2504.10374}}
  (\bibinfo{year}{2025}).
\newblock


\bibitem[Bhattacharyya et~al\mbox{.}(2026)]%
        {bhattacharyya2026social}
\bibfield{author}{\bibinfo{person}{Aanisha Bhattacharyya},
  \bibinfo{person}{Abhilekh Borah}, \bibinfo{person}{Yaman~Kumar Singla},
  \bibinfo{person}{Rajiv~Ratn Shah}, \bibinfo{person}{Changyou Chen}, {and}
  \bibinfo{person}{Balaji Krishnamurthy}.} \bibinfo{year}{2026}\natexlab{}.
\newblock \showarticletitle{Social agents: Collective intelligence improves llm
  predictions}. In \bibinfo{booktitle}{\emph{The Fourteenth International
  Conference on Learning Representations}}.
\newblock


\bibitem[Bothra et~al\mbox{.}(2024)]%
        {bothra2024lightweight}
\bibfield{author}{\bibinfo{person}{Rahul Bothra}, \bibinfo{person}{Venkat
  Arun}, \bibinfo{person}{Brighten Godfrey}, \bibinfo{person}{Akshay Narayan},
  {and} \bibinfo{person}{Ahmed Saeed}.} \bibinfo{year}{2024}\natexlab{}.
\newblock \showarticletitle{Lightweight Automated Reasoning for Network
  Architectures}. In \bibinfo{booktitle}{\emph{Proceedings of the 23rd ACM
  Workshop on Hot Topics in Networks}}. \bibinfo{pages}{237--245}.
\newblock


\bibitem[Bowman et~al\mbox{.}(2022)]%
        {bowman2022measuring}
\bibfield{author}{\bibinfo{person}{Samuel~R. Bowman}, \bibinfo{person}{Jeeyoon
  Hyun}, \bibinfo{person}{Ethan Perez}, \bibinfo{person}{Edwin Chen},
  \bibinfo{person}{Craig Pettit}, \bibinfo{person}{Scott Heiner},
  \bibinfo{person}{Kamil{\.e} Luko{\v{s}}i{\=u}t{\.e}}, \bibinfo{person}{Amanda
  Askell}, \bibinfo{person}{Andy Jones}, \bibinfo{person}{Anna Chen},
  \bibinfo{person}{Anna Goldie}, \bibinfo{person}{Azalia Mirhoseini},
  \bibinfo{person}{Cameron McKinnon}, \bibinfo{person}{Christopher Olah},
  \bibinfo{person}{Daniela Amodei}, \bibinfo{person}{Dario Amodei},
  \bibinfo{person}{Dawn Drain}, \bibinfo{person}{Dustin Li},
  \bibinfo{person}{Eli Tran-Johnson}, \bibinfo{person}{Jack Clark},
  \bibinfo{person}{Jackson Kernion}, \bibinfo{person}{Jamie Kerr},
  \bibinfo{person}{Jared Mueller}, \bibinfo{person}{Jeffrey Ladish},
  \bibinfo{person}{Joshua Landau}, \bibinfo{person}{Kamal Ndousse},
  \bibinfo{person}{Liane Lovitt}, \bibinfo{person}{Nelson Elhage},
  \bibinfo{person}{Nicholas Schiefer}, \bibinfo{person}{Nicholas Joseph},
  \bibinfo{person}{Noem{\'i} Mercado}, \bibinfo{person}{Nova DasSarma},
  \bibinfo{person}{Robin Larson}, \bibinfo{person}{Sam McCandlish},
  \bibinfo{person}{Sandipan Kundu}, \bibinfo{person}{Scott Johnston},
  \bibinfo{person}{Shauna Kravec}, \bibinfo{person}{Sheer El~Showk},
  \bibinfo{person}{Stanislav Fort}, \bibinfo{person}{Timothy Telleen-Lawton},
  \bibinfo{person}{Tom Brown}, \bibinfo{person}{Tom Henighan},
  \bibinfo{person}{Tristan Hume}, \bibinfo{person}{Yuntao Bai},
  \bibinfo{person}{Zac Hatfield-Dodds}, \bibinfo{person}{Ben Mann}, {and}
  \bibinfo{person}{Jared Kaplan}.} \bibinfo{year}{2022}\natexlab{}.
\newblock \showarticletitle{Measuring progress on scalable oversight for large
  language models}.
\newblock \bibinfo{journal}{\emph{arXiv preprint arXiv:2211.03540}}
  (\bibinfo{year}{2022}).
\newblock


\bibitem[Bullot et~al\mbox{.}(2008)]%
        {bullot2008situatedness}
\bibfield{author}{\bibinfo{person}{Thomas Bullot}, \bibinfo{person}{Rida
  Khatoun}, \bibinfo{person}{Louis Hugues}, \bibinfo{person}{Dominique
  Ga{\"\i}ti}, {and} \bibinfo{person}{Leila Merghem-Boulahia}.}
  \bibinfo{year}{2008}\natexlab{}.
\newblock \showarticletitle{A situatedness-based Knowledge Plane for autonomic
  networking}.
\newblock \bibinfo{journal}{\emph{International Journal of Network Management}}
  \bibinfo{volume}{18}, \bibinfo{number}{2} (\bibinfo{year}{2008}),
  \bibinfo{pages}{171--193}.
\newblock


\bibitem[Chan et~al\mbox{.}(2025)]%
        {chan2025infrastructure}
\bibfield{author}{\bibinfo{person}{Alan Chan}, \bibinfo{person}{Kevin Wei},
  \bibinfo{person}{Sihao Huang}, \bibinfo{person}{Nitarshan Rajkumar},
  \bibinfo{person}{Elija Perrier}, \bibinfo{person}{Seth Lazar},
  \bibinfo{person}{Gillian~K. Hadfield}, {and} \bibinfo{person}{Markus
  Anderljung}.} \bibinfo{year}{2025}\natexlab{}.
\newblock \showarticletitle{Infrastructure for AI agents}.
\newblock \bibinfo{journal}{\emph{arXiv preprint arXiv:2501.10114}}
  (\bibinfo{year}{2025}).
\newblock


\bibitem[Chen et~al\mbox{.}(2023)]%
        {chen2023frugalgpt}
\bibfield{author}{\bibinfo{person}{Lingjiao Chen}, \bibinfo{person}{Matei
  Zaharia}, {and} \bibinfo{person}{James Zou}.}
  \bibinfo{year}{2023}\natexlab{}.
\newblock \showarticletitle{{FrugalGPT}: How to Use Large Language Models While
  Reducing Cost and Improving Performance}.
\newblock \bibinfo{journal}{\emph{arXiv preprint arXiv:2305.05176}}
  (\bibinfo{year}{2023}).
\newblock
\urldef\tempurl%
\url{https://arxiv.org/abs/2305.05176}
\showURL{%
\tempurl}
\newblock
\shownote{Later published in Transactions on Machine Learning Research (TMLR),
  2024}.


\bibitem[Chen et~al\mbox{.}(2025)]%
        {chen2025harnessing}
\bibfield{author}{\bibinfo{person}{Zhijun Chen}, \bibinfo{person}{Xiaodong Lu},
  \bibinfo{person}{Jingzheng Li}, \bibinfo{person}{Pengpeng Chen},
  \bibinfo{person}{Zhuoran Li}, \bibinfo{person}{Kai Sun},
  \bibinfo{person}{Yuankai Luo}, \bibinfo{person}{Qianren Mao},
  \bibinfo{person}{Ming Li}, \bibinfo{person}{Likang Xiao},
  \bibinfo{person}{Dingqi Yang}, \bibinfo{person}{Xiao Huang},
  \bibinfo{person}{Yikun Ban}, \bibinfo{person}{Hailong Sun}, {and}
  \bibinfo{person}{Philip~S. Yu}.} \bibinfo{year}{2025}\natexlab{}.
\newblock \showarticletitle{Harnessing multiple large language models: A survey
  on llm ensemble}.
\newblock \bibinfo{journal}{\emph{arXiv preprint arXiv:2502.18036}}
  (\bibinfo{year}{2025}).
\newblock


\bibitem[Clark et~al\mbox{.}(2003)]%
        {clark2003knowledge}
\bibfield{author}{\bibinfo{person}{David~D Clark}, \bibinfo{person}{Craig
  Partridge}, \bibinfo{person}{J~Christopher Ramming}, {and}
  \bibinfo{person}{John~T Wroclawski}.} \bibinfo{year}{2003}\natexlab{}.
\newblock \showarticletitle{A knowledge plane for the internet}. In
  \bibinfo{booktitle}{\emph{Proceedings of the 2003 conference on Applications,
  technologies, architectures, and protocols for computer communications}}.
  \bibinfo{pages}{3--10}.
\newblock


\bibitem[Clemm et~al\mbox{.}(2022)]%
        {clemm2022intent}
\bibfield{author}{\bibinfo{person}{Alexander Clemm}, \bibinfo{person}{Laurent
  Ciavaglia}, \bibinfo{person}{Lisandro~Zambenedetti Granville}, {and}
  \bibinfo{person}{Jeff Tantsura}.} \bibinfo{year}{2022}\natexlab{}.
\newblock \bibinfo{booktitle}{\emph{Intent-{Based} {Networking} -- {Concepts}
  and {Definitions}}}.
\newblock \bibinfo{type}{Request for {Comments}} RFC 9315.
  \bibinfo{institution}{Internet Engineering Task Force}.
\newblock
\href{https://doi.org/10.17487/RFC9315}{doi:\nolinkurl{10.17487/RFC9315}}


\bibitem[{Coinbase Threat Intelligence}(2022)]%
        {coinbase2022celer}
\bibfield{author}{\bibinfo{person}{{Coinbase Threat Intelligence}}.}
  \bibinfo{year}{2022}\natexlab{}.
\newblock \bibinfo{booktitle}{\emph{Celer {Bridge} Incident Analysis}}.
\newblock The Coinbase Blog.
\newblock
\urldef\tempurl%
\url{https://www.coinbase.com/blog/celer-bridge-incident-analysis}
\showURL{%
Retrieved July 9, 2026 from \tempurl}


\bibitem[Debenedetti et~al\mbox{.}(2025)]%
        {debenedetti2025camel}
\bibfield{author}{\bibinfo{person}{Edoardo Debenedetti}, \bibinfo{person}{Ilia
  Shumailov}, \bibinfo{person}{Tianqi Fan}, \bibinfo{person}{Jamie Hayes},
  \bibinfo{person}{Nicholas Carlini}, \bibinfo{person}{Daniel Fabian},
  \bibinfo{person}{Christoph Kern}, \bibinfo{person}{Chongyang Shi},
  \bibinfo{person}{Andreas Terzis}, {and} \bibinfo{person}{Florian
  Tram{\`e}r}.} \bibinfo{year}{2025}\natexlab{}.
\newblock \showarticletitle{Defeating Prompt Injections by Design}.
\newblock \bibinfo{journal}{\emph{arXiv preprint arXiv:2503.18813}}
  (\bibinfo{year}{2025}).
\newblock
\urldef\tempurl%
\url{https://arxiv.org/abs/2503.18813}
\showURL{%
\tempurl}
\newblock
\shownote{Google DeepMind; capability-based custom interpreter separating
  control/data flow}.


\bibitem[Debenedetti et~al\mbox{.}(2024)]%
        {debenedetti2024agentdojo}
\bibfield{author}{\bibinfo{person}{Edoardo Debenedetti}, \bibinfo{person}{Jie
  Zhang}, \bibinfo{person}{Mislav Balunovic}, \bibinfo{person}{Luca
  Beurer-Kellner}, \bibinfo{person}{Marc Fischer}, {and}
  \bibinfo{person}{Florian Tram{\`e}r}.} \bibinfo{year}{2024}\natexlab{}.
\newblock \showarticletitle{Agentdojo: A dynamic environment to evaluate prompt
  injection attacks and defenses for llm agents}.
\newblock \bibinfo{journal}{\emph{Advances in Neural Information Processing
  Systems}}  \bibinfo{volume}{37} (\bibinfo{year}{2024}),
  \bibinfo{pages}{82895--82920}.
\newblock


\bibitem[Dhamdhere et~al\mbox{.}(2018)]%
        {dhamdhere2018congestion}
\bibfield{author}{\bibinfo{person}{Amogh Dhamdhere}, \bibinfo{person}{David~D.
  Clark}, \bibinfo{person}{Alexander Gamero-Garrido}, \bibinfo{person}{Matthew
  Luckie}, \bibinfo{person}{Ricky K.~P. Mok}, \bibinfo{person}{Gautam Akiwate},
  \bibinfo{person}{Kabir Gogia}, \bibinfo{person}{Vaibhav Bajpai},
  \bibinfo{person}{Alex~C. Snoeren}, {and} \bibinfo{person}{kc claffy}.}
  \bibinfo{year}{2018}\natexlab{}.
\newblock \showarticletitle{Inferring Persistent Interdomain Congestion}. In
  \bibinfo{booktitle}{\emph{Proceedings of the 2018 Conference of the ACM
  Special Interest Group on Data Communication (SIGCOMM '18)}}.
  \bibinfo{publisher}{ACM}, \bibinfo{address}{Budapest, Hungary},
  \bibinfo{pages}{1--15}.
\newblock
\href{https://doi.org/10.1145/3230543.3230549}{doi:\nolinkurl{10.1145/3230543.3230549}}


\bibitem[El-Hassany et~al\mbox{.}(2018)]%
        {el2018netcomplete}
\bibfield{author}{\bibinfo{person}{Ahmed El-Hassany}, \bibinfo{person}{Petar
  Tsankov}, \bibinfo{person}{Laurent Vanbever}, {and} \bibinfo{person}{Martin
  Vechev}.} \bibinfo{year}{2018}\natexlab{}.
\newblock \showarticletitle{$\{$NetComplete$\}$: Practical $\{$Network-Wide$\}$
  configuration synthesis with autocompletion}. In
  \bibinfo{booktitle}{\emph{15th USENIX Symposium on Networked Systems Design
  and Implementation (NSDI 18)}}. \bibinfo{pages}{579--594}.
\newblock


\bibitem[Feamster and Balakrishnan(2005)]%
        {feamster2005detecting}
\bibfield{author}{\bibinfo{person}{Nick Feamster} {and} \bibinfo{person}{Hari
  Balakrishnan}.} \bibinfo{year}{2005}\natexlab{}.
\newblock \showarticletitle{Detecting BGP configuration faults with static
  analysis}. In \bibinfo{booktitle}{\emph{Proceedings of the 2nd conference on
  Symposium on Networked Systems Design \& Implementation-Volume 2}}.
  \bibinfo{pages}{43--56}.
\newblock


\bibitem[Feamster and Rexford(2018)]%
        {self-driving-net}
\bibfield{author}{\bibinfo{person}{Nick Feamster} {and}
  \bibinfo{person}{Jennifer Rexford}.} \bibinfo{year}{2018}\natexlab{}.
\newblock \showarticletitle{Why (and How) Networks Should Run Themselves}. In
  \bibinfo{booktitle}{\emph{Proceedings of the 2018 Applied Networking Research
  Workshop}} (Montreal, QC, Canada) \emph{(\bibinfo{series}{ANRW '18})}.
  \bibinfo{publisher}{Association for Computing Machinery},
  \bibinfo{address}{New York, NY, USA}, \bibinfo{pages}{20}.
\newblock
\showISBNx{9781450355858}
\href{https://doi.org/10.1145/3232755.3234555}{doi:\nolinkurl{10.1145/3232755.3234555}}


\bibitem[Fogel et~al\mbox{.}(2015)]%
        {fogel2015general}
\bibfield{author}{\bibinfo{person}{Ari Fogel}, \bibinfo{person}{Stanley Fung},
  \bibinfo{person}{Luis Pedrosa}, \bibinfo{person}{Meg Walraed-Sullivan},
  \bibinfo{person}{Ramesh Govindan}, \bibinfo{person}{Ratul Mahajan}, {and}
  \bibinfo{person}{Todd Millstein}.} \bibinfo{year}{2015}\natexlab{}.
\newblock \showarticletitle{A general approach to network configuration
  analysis}. In \bibinfo{booktitle}{\emph{12th USENIX Symposium on Networked
  Systems Design and Implementation (NSDI 15)}}. \bibinfo{pages}{469--483}.
\newblock


\bibitem[Gaba et~al\mbox{.}(2026)]%
        {gaba2026benchmarking}
\bibfield{author}{\bibinfo{person}{Vinay Gaba}, \bibinfo{person}{Ankit Mathur},
  \bibinfo{person}{Rishabh Singh}, \bibinfo{person}{Patrick Wendell}, {and}
  \bibinfo{person}{Matei Zaharia}.} \bibinfo{year}{2026}\natexlab{}.
\newblock \bibinfo{title}{Benchmarking Coding Agents on {Databricks}'
  Multi-Million Line Codebase}.
\newblock \bibinfo{howpublished}{Databricks Blog}.
\newblock
\urldef\tempurl%
\url{https://www.databricks.com/blog/benchmarking-coding-agents-databricks-multi-million-line-codebase}
\showURL{%
\tempurl}
\newblock
\shownote{Accessed: 2026-07-16}.


\bibitem[Gao and Rexford(2001)]%
        {gao2001stable}
\bibfield{author}{\bibinfo{person}{Lixin Gao} {and} \bibinfo{person}{Jennifer
  Rexford}.} \bibinfo{year}{2001}\natexlab{}.
\newblock \showarticletitle{Stable Internet routing without global
  coordination}.
\newblock \bibinfo{journal}{\emph{IEEE/ACM Transactions on networking}}
  \bibinfo{volume}{9}, \bibinfo{number}{6} (\bibinfo{year}{2001}),
  \bibinfo{pages}{681--692}.
\newblock


\bibitem[Greenblatt et~al\mbox{.}(2024)]%
        {greenblatt2024aicontrol}
\bibfield{author}{\bibinfo{person}{Ryan Greenblatt}, \bibinfo{person}{Buck
  Shlegeris}, \bibinfo{person}{Kshitij Sachan}, {and} \bibinfo{person}{Fabien
  Roger}.} \bibinfo{year}{2024}\natexlab{}.
\newblock \showarticletitle{AI Control: Improving Safety Despite Intentional
  Subversion}.
\newblock \bibinfo{journal}{\emph{arXiv preprint arXiv:2312.06942}}
  (\bibinfo{year}{2024}).
\newblock
\urldef\tempurl%
\url{http://arxiv.org/abs/2312.06942}
\showURL{%
\tempurl}


\bibitem[Gupta et~al\mbox{.}(2018)]%
        {gupta2018sonata}
\bibfield{author}{\bibinfo{person}{Arpit Gupta}, \bibinfo{person}{Rob
  Harrison}, \bibinfo{person}{Marco Canini}, \bibinfo{person}{Nick Feamster},
  \bibinfo{person}{Jennifer Rexford}, {and} \bibinfo{person}{Walter
  Willinger}.} \bibinfo{year}{2018}\natexlab{}.
\newblock \showarticletitle{Sonata: Query-driven streaming network telemetry}.
  In \bibinfo{booktitle}{\emph{Proceedings of the 2018 conference of the ACM
  special interest group on data communication}}. \bibinfo{pages}{357--371}.
\newblock


\bibitem[Inan et~al\mbox{.}(2023)]%
        {inan2023llamaguard}
\bibfield{author}{\bibinfo{person}{Hakan Inan}, \bibinfo{person}{Kartikeya
  Upasani}, \bibinfo{person}{Jianfeng Chi}, \bibinfo{person}{Rashi Rungta},
  \bibinfo{person}{Krithika Iyer}, \bibinfo{person}{Yuning Mao},
  \bibinfo{person}{Michael Tontchev}, \bibinfo{person}{Qing Hu},
  \bibinfo{person}{Brian Fuller}, \bibinfo{person}{Davide Testuggine}, {and}
  \bibinfo{person}{Madian Khabsa}.} \bibinfo{year}{2023}\natexlab{}.
\newblock \showarticletitle{Llama Guard: LLM-based input-output safeguard for
  human-AI conversations}.
\newblock \bibinfo{journal}{\emph{arXiv preprint arXiv:2312.06674}}
  (\bibinfo{year}{2023}).
\newblock


\bibitem[Jacobs et~al\mbox{.}(2018)]%
        {jacobs2018refining}
\bibfield{author}{\bibinfo{person}{Arthur~Selle Jacobs},
  \bibinfo{person}{Ricardo~Jos{\'e} Pfitscher}, \bibinfo{person}{Ronaldo~Alves
  Ferreira}, {and} \bibinfo{person}{Lisandro~Zambenedetti Granville}.}
  \bibinfo{year}{2018}\natexlab{}.
\newblock \showarticletitle{Refining network intents for self-driving
  networks}. In \bibinfo{booktitle}{\emph{Proceedings of the Afternoon Workshop
  on Self-Driving Networks}}. \bibinfo{pages}{15--21}.
\newblock


\bibitem[Jacobs et~al\mbox{.}(2021)]%
        {jacobs2021hey}
\bibfield{author}{\bibinfo{person}{Arthur~S Jacobs}, \bibinfo{person}{Ricardo~J
  Pfitscher}, \bibinfo{person}{Rafael~H Ribeiro}, \bibinfo{person}{Ronaldo~A
  Ferreira}, \bibinfo{person}{Lisandro~Z Granville}, \bibinfo{person}{Walter
  Willinger}, {and} \bibinfo{person}{Sanjay~G Rao}.}
  \bibinfo{year}{2021}\natexlab{}.
\newblock \showarticletitle{Hey, lumi! using natural language for
  $\{$intent-based$\}$ network management}. In \bibinfo{booktitle}{\emph{2021
  usenix annual technical conference (usenix atc 21)}}.
  \bibinfo{pages}{625--639}.
\newblock


\bibitem[Jitkrittum et~al\mbox{.}(2023)]%
        {jitkrittum2023confidence}
\bibfield{author}{\bibinfo{person}{Wittawat Jitkrittum}, \bibinfo{person}{Neha
  Gupta}, \bibinfo{person}{Aditya~Krishna Menon}, \bibinfo{person}{Harikrishna
  Narasimhan}, \bibinfo{person}{Ankit~Singh Rawat}, {and}
  \bibinfo{person}{Sanjiv Kumar}.} \bibinfo{year}{2023}\natexlab{}.
\newblock \showarticletitle{When Does Confidence-Based Cascade Deferral
  Suffice?}. In \bibinfo{booktitle}{\emph{Advances in Neural Information
  Processing Systems}}, Vol.~\bibinfo{volume}{36}. \bibinfo{pages}{9891--9906}.
\newblock


\bibitem[Kale et~al\mbox{.}(2025)]%
        {kale2025reliable}
\bibfield{author}{\bibinfo{person}{Neil Kale}, \bibinfo{person}{Chen Bo~Calvin
  Zhang}, \bibinfo{person}{Kevin Zhu}, \bibinfo{person}{Ankit Aich},
  \bibinfo{person}{Paula Rodriguez}, \bibinfo{person}{Scale~Red Team},
  \bibinfo{person}{Christina~Q Knight}, {and} \bibinfo{person}{Zifan Wang}.}
  \bibinfo{year}{2025}\natexlab{}.
\newblock \showarticletitle{Reliable Weak-to-Strong Monitoring of LLM Agents}.
\newblock \bibinfo{journal}{\emph{arXiv preprint arXiv:2508.19461}}
  (\bibinfo{year}{2025}).
\newblock


\bibitem[Kang and Li(2024)]%
        {kang2024r2guard}
\bibfield{author}{\bibinfo{person}{Mintong Kang} {and} \bibinfo{person}{Bo
  Li}.} \bibinfo{year}{2024}\natexlab{}.
\newblock \showarticletitle{r2-guard: Robust reasoning enabled LLM guardrail
  via knowledge-enhanced logical reasoning}.
\newblock \bibinfo{journal}{\emph{arXiv preprint arXiv:2407.05557}}
  (\bibinfo{year}{2024}).
\newblock


\bibitem[Katz-Bassett et~al\mbox{.}(2008)]%
        {katzbassett2008hubble}
\bibfield{author}{\bibinfo{person}{Ethan Katz-Bassett},
  \bibinfo{person}{Harsha~V. Madhyastha}, \bibinfo{person}{John~P. John},
  \bibinfo{person}{Arvind Krishnamurthy}, \bibinfo{person}{David Wetherall},
  {and} \bibinfo{person}{Thomas Anderson}.} \bibinfo{year}{2008}\natexlab{}.
\newblock \showarticletitle{Studying Black Holes in the Internet with Hubble}.
  In \bibinfo{booktitle}{\emph{5th USENIX Symposium on Networked Systems Design
  and Implementation (NSDI 08)}}. \bibinfo{publisher}{USENIX Association},
  \bibinfo{address}{San Francisco, CA}.
\newblock


\bibitem[Kazemian et~al\mbox{.}(2012)]%
        {kazemian2012header}
\bibfield{author}{\bibinfo{person}{Peyman Kazemian}, \bibinfo{person}{George
  Varghese}, {and} \bibinfo{person}{Nick McKeown}.}
  \bibinfo{year}{2012}\natexlab{}.
\newblock \showarticletitle{Header space analysis: Static checking for
  networks}. In \bibinfo{booktitle}{\emph{9th USENIX Symposium on Networked
  Systems Design and Implementation (NSDI 12)}}. \bibinfo{pages}{113--126}.
\newblock


\bibitem[Kenton et~al\mbox{.}(2024)]%
        {kenton2024scalable}
\bibfield{author}{\bibinfo{person}{Zachary Kenton}, \bibinfo{person}{Noah~Y.
  Siegel}, \bibinfo{person}{J{\'a}nos Kram{\'a}r}, \bibinfo{person}{Jonah
  Brown-Cohen}, \bibinfo{person}{Samuel Albanie}, \bibinfo{person}{Jannis
  Bulian}, \bibinfo{person}{Rishabh Agarwal}, \bibinfo{person}{David Lindner},
  \bibinfo{person}{Yunhao Tang}, \bibinfo{person}{Noah~D. Goodman}, {and}
  \bibinfo{person}{Rohin Shah}.} \bibinfo{year}{2024}\natexlab{}.
\newblock \showarticletitle{On scalable oversight with weak LLMs judging strong
  LLMs}.
\newblock \bibinfo{journal}{\emph{Advances in Neural Information Processing
  Systems}}  \bibinfo{volume}{37} (\bibinfo{year}{2024}),
  \bibinfo{pages}{75229--75276}.
\newblock


\bibitem[Kfoury et~al\mbox{.}(2021)]%
        {kfoury2021exhaustive}
\bibfield{author}{\bibinfo{person}{Elie~F Kfoury}, \bibinfo{person}{Jorge
  Crichigno}, {and} \bibinfo{person}{Elias Bou-Harb}.}
  \bibinfo{year}{2021}\natexlab{}.
\newblock \showarticletitle{An exhaustive survey on p4 programmable data plane
  switches: Taxonomy, applications, challenges, and future trends}.
\newblock \bibinfo{journal}{\emph{IEEE access}}  \bibinfo{volume}{9}
  (\bibinfo{year}{2021}), \bibinfo{pages}{87094--87155}.
\newblock


\bibitem[Kim and Feamster(2013)]%
        {kim2013improving}
\bibfield{author}{\bibinfo{person}{Hyojoon Kim} {and} \bibinfo{person}{Nick
  Feamster}.} \bibinfo{year}{2013}\natexlab{}.
\newblock \showarticletitle{Improving network management with software defined
  networking}.
\newblock \bibinfo{journal}{\emph{IEEE Communications magazine}}
  \bibinfo{volume}{51}, \bibinfo{number}{2} (\bibinfo{year}{2013}),
  \bibinfo{pages}{114--119}.
\newblock


\bibitem[Kolawole et~al\mbox{.}(2024)]%
        {kolawole2024agreement}
\bibfield{author}{\bibinfo{person}{Steven Kolawole}, \bibinfo{person}{Don
  Dennis}, \bibinfo{person}{Ameet Talwalkar}, {and} \bibinfo{person}{Virginia
  Smith}.} \bibinfo{year}{2024}\natexlab{}.
\newblock \showarticletitle{Agreement-Based Cascading for Efficient Inference}.
\newblock \bibinfo{journal}{\emph{Transactions on Machine Learning Research}}
  (\bibinfo{year}{2024}).
\newblock
\urldef\tempurl%
\url{https://arxiv.org/abs/2407.02348}
\showURL{%
\tempurl}
\newblock
\shownote{arXiv:2407.02348}.


\bibitem[Kossmann et~al\mbox{.}(2024)]%
        {kossmann2024cascadeserve}
\bibfield{author}{\bibinfo{person}{Ferdi Kossmann}, \bibinfo{person}{Ziniu Wu},
  \bibinfo{person}{Alex Turk}, \bibinfo{person}{Nesime Tatbul},
  \bibinfo{person}{Lei Cao}, {and} \bibinfo{person}{Samuel Madden}.}
  \bibinfo{year}{2024}\natexlab{}.
\newblock \showarticletitle{{CascadeServe}: Unlocking Model Cascades for
  Inference Serving}.
\newblock \bibinfo{journal}{\emph{arXiv preprint arXiv:2406.14424}}
  (\bibinfo{year}{2024}).
\newblock
\urldef\tempurl%
\url{https://arxiv.org/abs/2406.14424}
\showURL{%
\tempurl}


\bibitem[Kutasov et~al\mbox{.}(2025)]%
        {kutasov2025shadearena}
\bibfield{author}{\bibinfo{person}{Jonathan Kutasov}, \bibinfo{person}{Yuqi
  Sun}, \bibinfo{person}{Paul Colognese}, \bibinfo{person}{Teun van~der Weij},
  \bibinfo{person}{Linda Petrini}, \bibinfo{person}{Chen Bo~Calvin Zhang},
  \bibinfo{person}{John Hughes}, \bibinfo{person}{Xiang Deng},
  \bibinfo{person}{Henry Sleight}, \bibinfo{person}{Tyler Tracy},
  \bibinfo{person}{Buck Shlegeris}, {and} \bibinfo{person}{Joe Benton}.}
  \bibinfo{year}{2025}\natexlab{}.
\newblock \bibinfo{title}{SHADE-Arena: Evaluating Sabotage and Monitoring in
  LLM Agents}.
\newblock
\showeprint[arxiv]{2506.15740}~[cs.AI]
\urldef\tempurl%
\url{https://arxiv.org/abs/2506.15740}
\showURL{%
\tempurl}


\bibitem[Kwon et~al\mbox{.}(2023)]%
        {kwon_efficient_2023}
\bibfield{author}{\bibinfo{person}{Woosuk Kwon}, \bibinfo{person}{Zhuohan Li},
  \bibinfo{person}{Siyuan Zhuang}, \bibinfo{person}{Ying Sheng},
  \bibinfo{person}{Lianmin Zheng}, \bibinfo{person}{Cody~Hao Yu},
  \bibinfo{person}{Joseph Gonzalez}, \bibinfo{person}{Hao Zhang}, {and}
  \bibinfo{person}{Ion Stoica}.} \bibinfo{year}{2023}\natexlab{}.
\newblock \showarticletitle{Efficient {Memory} {Management} for {Large}
  {Language} {Model} {Serving} with {PagedAttention}}. In
  \bibinfo{booktitle}{\emph{Proceedings of the 29th {Symposium} on {Operating}
  {Systems} {Principles}}} \emph{(\bibinfo{series}{{SOSP} '23})}.
  \bibinfo{publisher}{Association for Computing Machinery},
  \bibinfo{address}{New York, NY, USA}, \bibinfo{pages}{611--626}.
\newblock
\showISBNx{979-8-4007-0229-7}
\href{https://doi.org/10.1145/3600006.3613165}{doi:\nolinkurl{10.1145/3600006.3613165}}


\bibitem[Langley et~al\mbox{.}(2017)]%
        {langley2017quic}
\bibfield{author}{\bibinfo{person}{Adam Langley}, \bibinfo{person}{Alistair
  Riddoch}, \bibinfo{person}{Alyssa Wilk}, \bibinfo{person}{Antonio Vicente},
  \bibinfo{person}{Charles Krasic}, \bibinfo{person}{Dan Zhang},
  \bibinfo{person}{Fan Yang}, \bibinfo{person}{Fedor Kouranov},
  \bibinfo{person}{Ian Swett}, \bibinfo{person}{Janardhan Iyengar},
  \bibinfo{person}{Jeff Bailey}, \bibinfo{person}{Jeremy Dorfman},
  \bibinfo{person}{Jim Roskind}, \bibinfo{person}{Joanna Kulik},
  \bibinfo{person}{Patrik Westin}, \bibinfo{person}{Raman Tenneti},
  \bibinfo{person}{Robbie Shade}, \bibinfo{person}{Ryan Hamilton},
  \bibinfo{person}{Victor Vasiliev}, \bibinfo{person}{Wan-Teh Chang}, {and}
  \bibinfo{person}{Zhongyi Shi}.} \bibinfo{year}{2017}\natexlab{}.
\newblock \showarticletitle{The {QUIC} Transport Protocol: Design and
  Internet-Scale Deployment}. In \bibinfo{booktitle}{\emph{Proceedings of the
  Conference of the ACM Special Interest Group on Data Communication (SIGCOMM
  '17)}}. \bibinfo{publisher}{ACM}, \bibinfo{address}{Los Angeles, CA},
  \bibinfo{pages}{183--196}.
\newblock
\href{https://doi.org/10.1145/3098822.3098842}{doi:\nolinkurl{10.1145/3098822.3098842}}


\bibitem[Lantz et~al\mbox{.}(2010)]%
        {mininet}
\bibfield{author}{\bibinfo{person}{Bob Lantz}, \bibinfo{person}{Brandon
  Heller}, {and} \bibinfo{person}{Nick McKeown}.}
  \bibinfo{year}{2010}\natexlab{}.
\newblock \showarticletitle{A network in a laptop: rapid prototyping for
  software-defined networks}. In \bibinfo{booktitle}{\emph{Proceedings of the
  9th ACM SIGCOMM Workshop on Hot Topics in Networks}}. \bibinfo{pages}{1--6}.
\newblock


\bibitem[Lepinski and Kent(2012)]%
        {lepinski_rfc_nodate}
\bibfield{author}{\bibinfo{person}{Matt Lepinski} {and}
  \bibinfo{person}{Stephen Kent}.} \bibinfo{year}{2012}\natexlab{}.
\newblock \bibinfo{booktitle}{\emph{An {Infrastructure} to {Support} {Secure}
  {Internet} {Routing}}}.
\newblock \bibinfo{type}{Request for {Comments}} RFC 6480.
  \bibinfo{institution}{Internet Engineering Task Force}.
\newblock
\href{https://doi.org/10.17487/RFC6480}{doi:\nolinkurl{10.17487/RFC6480}}


\bibitem[Macedo et~al\mbox{.}(2008)]%
        {macedo2008knowledge}
\bibfield{author}{\bibinfo{person}{Daniel~F. Macedo}, \bibinfo{person}{Aldri~L.
  dos Santos}, \bibinfo{person}{Jos{\'e} Marcos~S. Nogueira}, {and}
  \bibinfo{person}{Guy Pujolle}.} \bibinfo{year}{2008}\natexlab{}.
\newblock \showarticletitle{A Knowledge Plane for Autonomic Context-Aware
  Wireless Mobile Ad Hoc Networks}. In \bibinfo{booktitle}{\emph{IFIP/IEEE
  International Conference on Management of Multimedia and Mobile Networks and
  Services (MMNS)}}. Springer, \bibinfo{pages}{1--13}.
\newblock


\bibitem[Madory(2022)]%
        {madory_what_2022}
\bibfield{author}{\bibinfo{person}{Doug Madory}.}
  \bibinfo{year}{2022}\natexlab{}.
\newblock \bibinfo{booktitle}{\emph{What can be learned from {BGP} hijacks
  targeting cryptocurrency services?}}
\newblock APNIC Blog.
\newblock
\urldef\tempurl%
\url{https://blog.apnic.net/2022/11/07/what-can-be-learned-from-bgp-hijacks-targeting-cryptocurrency-services/}
\showURL{%
Retrieved July 9, 2026 from \tempurl}


\bibitem[Mahajan et~al\mbox{.}(2002)]%
        {mahajan2002understanding}
\bibfield{author}{\bibinfo{person}{Ratul Mahajan}, \bibinfo{person}{David
  Wetherall}, {and} \bibinfo{person}{Tom Anderson}.}
  \bibinfo{year}{2002}\natexlab{}.
\newblock \showarticletitle{Understanding BGP misconfiguration}.
\newblock \bibinfo{journal}{\emph{ACM SIGCOMM Computer Communication Review}}
  \bibinfo{volume}{32}, \bibinfo{number}{4} (\bibinfo{year}{2002}),
  \bibinfo{pages}{3--16}.
\newblock


\bibitem[Mao et~al\mbox{.}(2016)]%
        {mao2016resource}
\bibfield{author}{\bibinfo{person}{Hongzi Mao}, \bibinfo{person}{Mohammad
  Alizadeh}, \bibinfo{person}{Ishai Menache}, {and} \bibinfo{person}{Srikanth
  Kandula}.} \bibinfo{year}{2016}\natexlab{}.
\newblock \showarticletitle{Resource management with deep reinforcement
  learning}. In \bibinfo{booktitle}{\emph{Proceedings of the 15th ACM workshop
  on hot topics in networks}}. \bibinfo{pages}{50--56}.
\newblock


\bibitem[Mao et~al\mbox{.}(2017)]%
        {mao2017neural}
\bibfield{author}{\bibinfo{person}{Hongzi Mao}, \bibinfo{person}{Ravi
  Netravali}, {and} \bibinfo{person}{Mohammad Alizadeh}.}
  \bibinfo{year}{2017}\natexlab{}.
\newblock \showarticletitle{Neural adaptive video streaming with pensieve}. In
  \bibinfo{booktitle}{\emph{Proceedings of the conference of the ACM special
  interest group on data communication}}. \bibinfo{pages}{197--210}.
\newblock


\bibitem[Mao et~al\mbox{.}(2019)]%
        {mao2019learning}
\bibfield{author}{\bibinfo{person}{Hongzi Mao}, \bibinfo{person}{Malte
  Schwarzkopf}, \bibinfo{person}{Shaileshh~Bojja Venkatakrishnan},
  \bibinfo{person}{Zili Meng}, {and} \bibinfo{person}{Mohammad Alizadeh}.}
  \bibinfo{year}{2019}\natexlab{}.
\newblock \showarticletitle{Learning scheduling algorithms for data processing
  clusters}.
\newblock In \bibinfo{booktitle}{\emph{Proceedings of the ACM special interest
  group on data communication}}. \bibinfo{pages}{270--288}.
\newblock


\bibitem[McPherson et~al\mbox{.}(2015)]%
        {mcpherson_considerations_2015}
\bibfield{author}{\bibinfo{person}{Danny~R. McPherson}, \bibinfo{person}{Shane
  Amante}, \bibinfo{person}{Eric Osterweil}, \bibinfo{person}{Larry Blunk},
  {and} \bibinfo{person}{Dave Mitchell}.} \bibinfo{year}{2015}\natexlab{}.
\newblock \bibinfo{booktitle}{\emph{Considerations for {Internet} {Routing}
  {Registries} ({IRRs}) and {Routing} {Policy} {Configuration}}}.
\newblock \bibinfo{type}{Request for {Comments}} RFC 7682.
  \bibinfo{institution}{Internet Engineering Task Force}.
\newblock
\href{https://doi.org/10.17487/RFC7682}{doi:\nolinkurl{10.17487/RFC7682}}


\bibitem[Mekrache et~al\mbox{.}(2024)]%
        {mekrache2024intentcentric}
\bibfield{author}{\bibinfo{person}{Adlen Mekrache}, \bibinfo{person}{Adlen
  Ksentini}, {and} \bibinfo{person}{Christos Verikoukis}.}
  \bibinfo{year}{2024}\natexlab{}.
\newblock \showarticletitle{Intent-Based Management of Next-Generation
  Networks: An {LLM}-Centric Approach}.
\newblock \bibinfo{journal}{\emph{IEEE Network}} \bibinfo{volume}{38},
  \bibinfo{number}{5} (\bibinfo{year}{2024}).
\newblock


\bibitem[Mestres et~al\mbox{.}(2017)]%
        {mestres2017knowledge}
\bibfield{author}{\bibinfo{person}{Albert Mestres}, \bibinfo{person}{Alberto
  Rodriguez-Natal}, \bibinfo{person}{Josep Carner}, \bibinfo{person}{Pere
  Barlet-Ros}, \bibinfo{person}{Eduard Alarc{\'o}n}, \bibinfo{person}{Marc
  Sol{\'e}}, \bibinfo{person}{Victor Munt{\'e}s-Mulero}, \bibinfo{person}{David
  Meyer}, \bibinfo{person}{Sharon Barkai}, \bibinfo{person}{Mike~J Hibbett},
  \bibinfo{person}{Giovani Estrada}, \bibinfo{person}{Khaldun Ma'ruf},
  \bibinfo{person}{Florin Coras}, \bibinfo{person}{Vina Ermagan},
  \bibinfo{person}{Hugo Latapie}, \bibinfo{person}{Chris Cassar},
  \bibinfo{person}{John Evans}, \bibinfo{person}{Fabio Maino},
  \bibinfo{person}{Jean Walrand}, {and} \bibinfo{person}{Albert Cabellos}.}
  \bibinfo{year}{2017}\natexlab{}.
\newblock \showarticletitle{Knowledge-defined networking}.
\newblock \bibinfo{journal}{\emph{ACM SIGCOMM Computer Communication Review}}
  \bibinfo{volume}{47}, \bibinfo{number}{3} (\bibinfo{year}{2017}),
  \bibinfo{pages}{2--10}.
\newblock


\bibitem[{Omnigent}(2026)]%
        {omnigent_policies_overview}
\bibfield{author}{\bibinfo{person}{{Omnigent}}.}
  \bibinfo{year}{2026}\natexlab{}.
\newblock \bibinfo{title}{Contextual Policies}.
\newblock \bibinfo{howpublished}{Omnigent Documentation}.
\newblock
\urldef\tempurl%
\url{https://omnigent.ai/docs/policies/overview}
\showURL{%
\tempurl}
\newblock
\shownote{Accessed: 2026-07-16}.


\bibitem[Ong et~al\mbox{.}(2024)]%
        {ong2024routellm}
\bibfield{author}{\bibinfo{person}{Isaac Ong}, \bibinfo{person}{Amjad
  Almahairi}, \bibinfo{person}{Vincent Wu}, \bibinfo{person}{Wei-Lin Chiang},
  \bibinfo{person}{Tianhao Wu}, \bibinfo{person}{Joseph~E. Gonzalez},
  \bibinfo{person}{M.~Waleed Kadous}, {and} \bibinfo{person}{Ion Stoica}.}
  \bibinfo{year}{2024}\natexlab{}.
\newblock \showarticletitle{{RouteLLM}: Learning to Route {LLMs} with
  Preference Data}.
\newblock \bibinfo{journal}{\emph{arXiv preprint arXiv:2406.18665}}
  (\bibinfo{year}{2024}).
\newblock
\urldef\tempurl%
\url{https://arxiv.org/abs/2406.18665}
\showURL{%
\tempurl}


\bibitem[Pacchiardi et~al\mbox{.}(2023)]%
        {pacchiardi2023liar}
\bibfield{author}{\bibinfo{person}{Lorenzo Pacchiardi},
  \bibinfo{person}{Alex~J. Chan}, \bibinfo{person}{S{\"o}ren Mindermann},
  \bibinfo{person}{Ilan Moscovitz}, \bibinfo{person}{Alexa~Y. Pan},
  \bibinfo{person}{Yarin Gal}, \bibinfo{person}{Owain Evans}, {and}
  \bibinfo{person}{Jan Brauner}.} \bibinfo{year}{2023}\natexlab{}.
\newblock \showarticletitle{How to Catch an AI Liar: Lie Detection in Black-Box
  LLMs by Asking Unrelated Questions}.
\newblock \bibinfo{journal}{\emph{arXiv preprint arXiv:2309.15840}}
  (\bibinfo{year}{2023}).
\newblock
\urldef\tempurl%
\url{http://arxiv.org/abs/2309.15840}
\showURL{%
\tempurl}


\bibitem[Raiciu et~al\mbox{.}(2012)]%
        {raiciu2012mptcp}
\bibfield{author}{\bibinfo{person}{Costin Raiciu}, \bibinfo{person}{Christoph
  Paasch}, \bibinfo{person}{Sebastien Barre}, \bibinfo{person}{Alan Ford},
  \bibinfo{person}{Michio Honda}, \bibinfo{person}{Fabien Duchene},
  \bibinfo{person}{Olivier Bonaventure}, {and} \bibinfo{person}{Mark Handley}.}
  \bibinfo{year}{2012}\natexlab{}.
\newblock \showarticletitle{How Hard Can It Be? Designing and Implementing a
  Deployable Multipath {TCP}}. In \bibinfo{booktitle}{\emph{9th USENIX
  Symposium on Networked Systems Design and Implementation (NSDI 12)}}.
  \bibinfo{publisher}{USENIX Association}, \bibinfo{address}{San Jose, CA},
  \bibinfo{pages}{399--412}.
\newblock


\bibitem[Ramanathan et~al\mbox{.}(2023)]%
        {ramanathan2023practical}
\bibfield{author}{\bibinfo{person}{Sivaramakrishnan Ramanathan},
  \bibinfo{person}{Ying Zhang}, \bibinfo{person}{Mohab Gawish},
  \bibinfo{person}{Yogesh Mundada}, \bibinfo{person}{Zhaodong Wang},
  \bibinfo{person}{Sangki Yun}, \bibinfo{person}{Eric Lippert},
  \bibinfo{person}{Walid Taha}, \bibinfo{person}{Minlan Yu}, {and}
  \bibinfo{person}{Jelena Mirkovic}.} \bibinfo{year}{2023}\natexlab{}.
\newblock \showarticletitle{Practical intent-driven routing configuration
  synthesis}. In \bibinfo{booktitle}{\emph{20th USENIX Symposium on Networked
  Systems Design and Implementation (NSDI 23)}}. \bibinfo{pages}{629--644}.
\newblock


\bibitem[Rebedea et~al\mbox{.}(2023)]%
        {rebedea2023nemo}
\bibfield{author}{\bibinfo{person}{Traian Rebedea}, \bibinfo{person}{Razvan
  Dinu}, \bibinfo{person}{Makesh~Narsimhan Sreedhar},
  \bibinfo{person}{Christopher Parisien}, {and} \bibinfo{person}{Jonathan
  Cohen}.} \bibinfo{year}{2023}\natexlab{}.
\newblock \showarticletitle{Nemo guardrails: A toolkit for controllable and
  safe llm applications with programmable rails}. In
  \bibinfo{booktitle}{\emph{Proceedings of the 2023 conference on empirical
  methods in natural language processing: system demonstrations}}.
  \bibinfo{pages}{431--445}.
\newblock


\bibitem[Roy et~al\mbox{.}(2024)]%
        {roy2024exploring}
\bibfield{author}{\bibinfo{person}{Devjeet Roy}, \bibinfo{person}{Xuchao
  Zhang}, \bibinfo{person}{Rashi Bhave}, \bibinfo{person}{Chetan Bansal},
  \bibinfo{person}{Pedro Las-Casas}, \bibinfo{person}{Rodrigo Fonseca}, {and}
  \bibinfo{person}{Saravan Rajmohan}.} \bibinfo{year}{2024}\natexlab{}.
\newblock \showarticletitle{Exploring {LLM}-Based Agents for Root Cause
  Analysis}. In \bibinfo{booktitle}{\emph{Companion Proceedings of the 32nd ACM
  International Conference on the Foundations of Software Engineering (FSE)}}.
  \bibinfo{pages}{208--219}.
\newblock


\bibitem[Saeed et~al\mbox{.}(2019)]%
        {saeed2019eiffel}
\bibfield{author}{\bibinfo{person}{Ahmed Saeed}, \bibinfo{person}{Yimeng Zhao},
  \bibinfo{person}{Nandita Dukkipati}, \bibinfo{person}{Ellen Zegura},
  \bibinfo{person}{Mostafa Ammar}, \bibinfo{person}{Khaled Harras}, {and}
  \bibinfo{person}{Amin Vahdat}.} \bibinfo{year}{2019}\natexlab{}.
\newblock \showarticletitle{Eiffel: Efficient and flexible software packet
  scheduling}. In \bibinfo{booktitle}{\emph{16th USENIX Symposium on Networked
  Systems Design and Implementation (NSDI 19)}}. \bibinfo{pages}{17--32}.
\newblock


\bibitem[Sivaraman et~al\mbox{.}(2016)]%
        {sivaraman2016programmable}
\bibfield{author}{\bibinfo{person}{Anirudh Sivaraman}, \bibinfo{person}{Suvinay
  Subramanian}, \bibinfo{person}{Mohammad Alizadeh}, \bibinfo{person}{Sharad
  Chole}, \bibinfo{person}{Shang-Tse Chuang}, \bibinfo{person}{Anurag Agrawal},
  \bibinfo{person}{Hari Balakrishnan}, \bibinfo{person}{Tom Edsall},
  \bibinfo{person}{Sachin Katti}, {and} \bibinfo{person}{Nick McKeown}.}
  \bibinfo{year}{2016}\natexlab{}.
\newblock \showarticletitle{Programmable packet scheduling at line rate}. In
  \bibinfo{booktitle}{\emph{Proceedings of the 2016 ACM SIGCOMM Conference}}.
  \bibinfo{pages}{44--57}.
\newblock


\bibitem[Soiffer et~al\mbox{.}(2025)]%
        {soiffer2025semantic}
\bibfield{author}{\bibinfo{person}{Duncan Soiffer}, \bibinfo{person}{Steven
  Kolawole}, {and} \bibinfo{person}{Virginia Smith}.}
  \bibinfo{year}{2025}\natexlab{}.
\newblock \showarticletitle{Semantic Agreement Enables Efficient Open-Ended
  {LLM} Cascades}. In \bibinfo{booktitle}{\emph{Proceedings of the 2025
  Conference on Empirical Methods in Natural Language Processing: Industry
  Track}}. \bibinfo{pages}{2499--2537}.
\newblock


\bibitem[Spring et~al\mbox{.}(2003)]%
        {spring2003pathinflation}
\bibfield{author}{\bibinfo{person}{Neil Spring}, \bibinfo{person}{Ratul
  Mahajan}, {and} \bibinfo{person}{Thomas Anderson}.}
  \bibinfo{year}{2003}\natexlab{}.
\newblock \showarticletitle{Quantifying the Causes of Path Inflation}. In
  \bibinfo{booktitle}{\emph{Proceedings of the 2003 Conference on Applications,
  Technologies, Architectures, and Protocols for Computer Communications
  (SIGCOMM '03)}}. \bibinfo{publisher}{ACM}, \bibinfo{address}{Karlsruhe,
  Germany}, \bibinfo{pages}{113--124}.
\newblock
\href{https://doi.org/10.1145/863955.863970}{doi:\nolinkurl{10.1145/863955.863970}}


\bibitem[Staufer et~al\mbox{.}(2026)]%
        {staufer20262025}
\bibfield{author}{\bibinfo{person}{Leon Staufer}, \bibinfo{person}{Kevin Feng},
  \bibinfo{person}{Kevin Wei}, \bibinfo{person}{Luke Bailey},
  \bibinfo{person}{Yawen Duan}, \bibinfo{person}{Mick Yang},
  \bibinfo{person}{A~Pinar Ozisik}, \bibinfo{person}{Stephen Casper}, {and}
  \bibinfo{person}{Noam Kolt}.} \bibinfo{year}{2026}\natexlab{}.
\newblock \showarticletitle{The 2025 ai agent index: Documenting technical and
  safety features of deployed agentic ai systems}. In
  \bibinfo{booktitle}{\emph{The 2026 ACM Conference on Fairness,
  Accountability, and Transparency}}. \bibinfo{pages}{1536--1576}.
\newblock


\bibitem[Swamy et~al\mbox{.}(2022)]%
        {swamy2022taurus}
\bibfield{author}{\bibinfo{person}{Tushar Swamy}, \bibinfo{person}{Alexander
  Rucker}, \bibinfo{person}{Muhammad Shahbaz}, \bibinfo{person}{Ishan Gaur},
  {and} \bibinfo{person}{Kunle Olukotun}.} \bibinfo{year}{2022}\natexlab{}.
\newblock \showarticletitle{Taurus: a data plane architecture for per-packet
  ML}. In \bibinfo{booktitle}{\emph{Proceedings of the 27th ACM International
  Conference on Architectural Support for Programming Languages and Operating
  Systems}}. \bibinfo{pages}{1099--1114}.
\newblock


\bibitem[{Taalas}(2026)]%
        {taalas2026company}
\bibfield{author}{\bibinfo{person}{{Taalas}}.} \bibinfo{year}{2026}\natexlab{}.
\newblock \bibinfo{title}{Taalas}.
\newblock
\urldef\tempurl%
\url{https://taalas.com/}
\showURL{%
\tempurl}
\newblock
\shownote{Accessed: 2026-06-13}.


\bibitem[Tian et~al\mbox{.}(2019)]%
        {tian2019safely}
\bibfield{author}{\bibinfo{person}{Bingchuan Tian}, \bibinfo{person}{Xinyi
  Zhang}, \bibinfo{person}{Ennan Zhai}, \bibinfo{person}{Hongqiang~Harry Liu},
  \bibinfo{person}{Qiaobo Ye}, \bibinfo{person}{Chunsheng Wang},
  \bibinfo{person}{Xin Wu}, \bibinfo{person}{Zhiming Ji},
  \bibinfo{person}{Yihong Sang}, \bibinfo{person}{Ming Zhang},
  \bibinfo{person}{Da Yu}, \bibinfo{person}{Chen Tian}, \bibinfo{person}{Haitao
  Zheng}, {and} \bibinfo{person}{Ben~Y. Zhao}.}
  \bibinfo{year}{2019}\natexlab{}.
\newblock \showarticletitle{Safely and automatically updating in-network ACL
  configurations with intent language}.
\newblock In \bibinfo{booktitle}{\emph{Proceedings of the ACM Special Interest
  Group on Data Communication}}. \bibinfo{pages}{214--226}.
\newblock


\bibitem[Tseng et~al\mbox{.}(2026)]%
        {sade2026}
\bibfield{author}{\bibinfo{person}{Kuan-Hao Tseng}, \bibinfo{person}{Niruth
  Bogahawatta}, \bibinfo{person}{Yasod Ginige}, \bibinfo{person}{Kosta Dekic},
  \bibinfo{person}{Arunan Sivanathan}, {and} \bibinfo{person}{Suranga
  Seneviratne}.} \bibinfo{year}{2026}\natexlab{}.
\newblock \showarticletitle{SADE: Symptom-Aware Diagnostic Escalation for
  LLM-Based Network Troubleshooting}.
\newblock \bibinfo{journal}{\emph{arXiv preprint arXiv:2605.04530}}
  (\bibinfo{year}{2026}).
\newblock


\bibitem[Usama et~al\mbox{.}(2019)]%
        {usama2019unsupervised}
\bibfield{author}{\bibinfo{person}{Muhammad Usama}, \bibinfo{person}{Junaid
  Qadir}, \bibinfo{person}{Aunn Raza}, \bibinfo{person}{Hunain Arif},
  \bibinfo{person}{Kok-Lim~Alvin Yau}, \bibinfo{person}{Yehia Elkhatib},
  \bibinfo{person}{Amir Hussain}, {and} \bibinfo{person}{Ala Al-Fuqaha}.}
  \bibinfo{year}{2019}\natexlab{}.
\newblock \showarticletitle{Unsupervised machine learning for networking:
  Techniques, applications and research challenges}.
\newblock \bibinfo{journal}{\emph{IEEE access}}  \bibinfo{volume}{7}
  (\bibinfo{year}{2019}), \bibinfo{pages}{65579--65615}.
\newblock


\bibitem[Wang et~al\mbox{.}(2025c)]%
        {wang2025bian}
\bibfield{author}{\bibinfo{person}{Chenxu Wang}, \bibinfo{person}{Xumiao
  Zhang}, \bibinfo{person}{Runwei Lu}, \bibinfo{person}{Xianshang Lin},
  \bibinfo{person}{Xuan Zeng}, \bibinfo{person}{Xinlei Zhang},
  \bibinfo{person}{Zhe An}, \bibinfo{person}{Gongwei Wu},
  \bibinfo{person}{Jiaqi Gao}, \bibinfo{person}{Chen Tian},
  \bibinfo{person}{Guihai Chen}, \bibinfo{person}{Guyue Liu},
  \bibinfo{person}{Yuhong Liao}, \bibinfo{person}{Tao Lin},
  \bibinfo{person}{Dennis Cai}, {and} \bibinfo{person}{Ennan Zhai}.}
  \bibinfo{year}{2025}\natexlab{c}.
\newblock \showarticletitle{Towards {LLM}-Based Failure Localization in
  Production-Scale Networks}. In \bibinfo{booktitle}{\emph{Proceedings of the
  ACM SIGCOMM 2025 Conference}}. \bibinfo{pages}{496--511}.
\newblock
\href{https://doi.org/10.1145/3718958.3750505}{doi:\nolinkurl{10.1145/3718958.3750505}}


\bibitem[Wang et~al\mbox{.}(2024a)]%
        {wang2024netassistant}
\bibfield{author}{\bibinfo{person}{Haopei Wang}, \bibinfo{person}{Anubhavnidhi
  Abhashkumar}, \bibinfo{person}{Changyu Lin}, \bibinfo{person}{Tianrong
  Zhang}, \bibinfo{person}{Xiaoming Gu}, \bibinfo{person}{Ning Ma},
  \bibinfo{person}{Chang Wu}, \bibinfo{person}{Songlin Liu},
  \bibinfo{person}{Wei Zhou}, \bibinfo{person}{Yongbin Dong},
  \bibinfo{person}{Weirong Jiang}, {and} \bibinfo{person}{Yi Wang}.}
  \bibinfo{year}{2024}\natexlab{a}.
\newblock \showarticletitle{{NetAssistant}: Dialogue Based Network Diagnosis in
  Data Center Networks}. In \bibinfo{booktitle}{\emph{21st USENIX Symposium on
  Networked Systems Design and Implementation (NSDI 24)}}.
  \bibinfo{pages}{2011--2024}.
\newblock


\bibitem[Wang et~al\mbox{.}(2025a)]%
        {networkarena2025}
\bibfield{author}{\bibinfo{person}{Zhihao Wang}, \bibinfo{person}{Alessandro
  Cornacchia}, \bibinfo{person}{Alessio Sacco}, \bibinfo{person}{Franco
  Galante}, \bibinfo{person}{Marco Canini}, {and} \bibinfo{person}{Dingde
  Jiang}.} \bibinfo{year}{2025}\natexlab{a}.
\newblock \showarticletitle{A Network Arena for Benchmarking AI Agents on
  Network Troubleshooting}.
\newblock \bibinfo{journal}{\emph{arXiv preprint arXiv:2512.16381}}
  (\bibinfo{year}{2025}).
\newblock


\bibitem[Wang et~al\mbox{.}(2025b)]%
        {wang2025confucius}
\bibfield{author}{\bibinfo{person}{Zhaodong Wang}, \bibinfo{person}{Samuel
  Lin}, \bibinfo{person}{Guanqing Yan}, \bibinfo{person}{Soudeh Ghorbani},
  \bibinfo{person}{Minlan Yu}, \bibinfo{person}{Jiawei Zhou},
  \bibinfo{person}{Nathan Hu}, \bibinfo{person}{Lopa Baruah},
  \bibinfo{person}{Sam Peters}, \bibinfo{person}{Srikanth Kamath},
  \bibinfo{person}{Jerry Yang}, {and} \bibinfo{person}{Ying Zhang}.}
  \bibinfo{year}{2025}\natexlab{b}.
\newblock \showarticletitle{Intent-Driven Network Management with Multi-Agent
  {LLMs}: The {Confucius} Framework}. In \bibinfo{booktitle}{\emph{Proceedings
  of the ACM SIGCOMM 2025 Conference}}. \bibinfo{pages}{347--362}.
\newblock
\href{https://doi.org/10.1145/3718958.3750537}{doi:\nolinkurl{10.1145/3718958.3750537}}


\bibitem[Wang et~al\mbox{.}(2024b)]%
        {wang2024rcagent}
\bibfield{author}{\bibinfo{person}{Zefan Wang}, \bibinfo{person}{Zichuan Liu},
  \bibinfo{person}{Yingying Zhang}, \bibinfo{person}{Aoxiao Zhong},
  \bibinfo{person}{Jihong Wang}, \bibinfo{person}{Fengbin Yin},
  \bibinfo{person}{Lunting Fan}, \bibinfo{person}{Lingfei Wu}, {and}
  \bibinfo{person}{Qingsong Wen}.} \bibinfo{year}{2024}\natexlab{b}.
\newblock \showarticletitle{{RCAgent}: Cloud Root Cause Analysis by Autonomous
  Agents with Tool-Augmented Large Language Models}. In
  \bibinfo{booktitle}{\emph{Proceedings of the 33rd ACM International
  Conference on Information and Knowledge Management (CIKM '24)}}.
  \bibinfo{pages}{4966--4974}.
\newblock


\bibitem[Wijesekara and Gunawardena(2023)]%
        {wijesekara2023comprehensive}
\bibfield{author}{\bibinfo{person}{Patikiri Arachchige Don Shehan~Nilmantha
  Wijesekara} {and} \bibinfo{person}{Subodha Gunawardena}.}
  \bibinfo{year}{2023}\natexlab{}.
\newblock \showarticletitle{A comprehensive survey on knowledge-defined
  networking}. In \bibinfo{booktitle}{\emph{Telecom}},
  Vol.~\bibinfo{volume}{4}. MDPI, \bibinfo{pages}{477--596}.
\newblock


\bibitem[Wiselman(2026)]%
        {wiselman2026memorysafety}
\bibfield{author}{\bibinfo{person}{Rayne Wiselman}.}
  \bibinfo{year}{2026}\natexlab{}.
\newblock \bibinfo{title}{Manage AI Memory Safety in Agentic Systems}.
\newblock \bibinfo{howpublished}{Microsoft Learn}.
\newblock
\urldef\tempurl%
\url{https://learn.microsoft.com/en-us/security/zero-trust/sfi/manage-agentic-memory-safety}
\showURL{%
\tempurl}
\newblock
\shownote{Accessed: 2026-07-16}.


\bibitem[Wu et~al\mbox{.}(2024)]%
        {wu2024fsecure}
\bibfield{author}{\bibinfo{person}{Fangzhou Wu}, \bibinfo{person}{Ethan
  Cecchetti}, {and} \bibinfo{person}{Chaowei Xiao}.}
  \bibinfo{year}{2024}\natexlab{}.
\newblock \showarticletitle{System-level defense against indirect prompt
  injection attacks: An information flow control perspective}.
\newblock \bibinfo{journal}{\emph{arXiv preprint arXiv:2409.19091}}
  (\bibinfo{year}{2024}).
\newblock
\urldef\tempurl%
\url{https://arxiv.org/abs/2409.19091}
\showURL{%
\tempurl}


\bibitem[Wu et~al\mbox{.}(2025)]%
        {tnautorca2025}
\bibfield{author}{\bibinfo{person}{Keyu Wu}, \bibinfo{person}{Qianjin Yu},
  \bibinfo{person}{Manlin Mei}, \bibinfo{person}{Ruiting Liu},
  \bibinfo{person}{Jun Wang}, \bibinfo{person}{Kailai Zhang},
  \bibinfo{person}{Yelun Bao}, \bibinfo{person}{Wenbo Wan},
  \bibinfo{person}{Rui Ye}, \bibinfo{person}{Bin He}, \bibinfo{person}{Junfeng
  Liao}, \bibinfo{person}{Yongsheng Du}, \bibinfo{person}{Zhiguo Yang},
  \bibinfo{person}{Kunlin Liu}, \bibinfo{person}{Lingjun Huang},
  \bibinfo{person}{Zhiwei Song}, \bibinfo{person}{YanQin Gao},
  \bibinfo{person}{Fang Tan}, \bibinfo{person}{Jin Yang}, {and}
  \bibinfo{person}{Ninglun Gu}.} \bibinfo{year}{2025}\natexlab{}.
\newblock \showarticletitle{{TN-AutoRCA}: Benchmark Construction and Agentic
  Framework for Self-Improving Alarm-Based Root Cause Analysis in
  Telecommunication Networks}.
\newblock \bibinfo{journal}{\emph{arXiv preprint arXiv:2507.18190}}
  (\bibinfo{year}{2025}).
\newblock


\bibitem[Xiang et~al\mbox{.}(2024)]%
        {xiang2024guardagent}
\bibfield{author}{\bibinfo{person}{Zhen Xiang}, \bibinfo{person}{Linzhi Zheng},
  \bibinfo{person}{Yanjie Li}, \bibinfo{person}{Junyuan Hong},
  \bibinfo{person}{Qinbin Li}, \bibinfo{person}{Han Xie},
  \bibinfo{person}{Jiawei Zhang}, \bibinfo{person}{Zidi Xiong},
  \bibinfo{person}{Chulin Xie}, \bibinfo{person}{Carl Yang},
  \bibinfo{person}{Dawn Song}, {and} \bibinfo{person}{Bo Li}.}
  \bibinfo{year}{2024}\natexlab{}.
\newblock \showarticletitle{GuardAgent: Safeguard LLM agents by a guard agent
  via knowledge-enabled reasoning}.
\newblock \bibinfo{journal}{\emph{arXiv preprint arXiv:2406.09187}}
  (\bibinfo{year}{2024}).
\newblock


\bibitem[Xu et~al\mbox{.}(2025)]%
        {xu2025openrca}
\bibfield{author}{\bibinfo{person}{Jinyang Xu}, \bibinfo{person}{Qiuai Zhang},
  \bibinfo{person}{Ziyi Zhong}, \bibinfo{person}{Shilin He},
  \bibinfo{person}{Chaoyun Zhang}, \bibinfo{person}{Qingwei Lin},
  \bibinfo{person}{Dan Pei}, \bibinfo{person}{Pinjia He},
  \bibinfo{person}{Dongmei Zhang}, {and} \bibinfo{person}{Qi Zhang}.}
  \bibinfo{year}{2025}\natexlab{}.
\newblock \showarticletitle{{OpenRCA}: Can Large Language Models Locate the
  Root Cause of Software Failures?}. In \bibinfo{booktitle}{\emph{International
  Conference on Learning Representations (ICLR 2025)}}.
\newblock
\newblock
\shownote{Poster}.


\end{thebibliography}
